# Software engineering for artificial intelligence and machine learning software: A systematic literature review

Elizamary Nascimento[1*], Anh Nguyen-Duc[2], Ingrid Sundbø[2], Tayana Conte[1]

[1] Institute of Computing, Universidade Federal do Amazonas, Manaus, Amazonas, Brasil
[2] School of Business, University of South-Eastern Norway, Bø i Telemark, Norway
[*]elizamary.souza@icomp.ufam.edu.br

**Abstract:** Artificial Intelligence (AI) or Machine Learning (ML) systems have been widely adopted as value propositions by companies in all industries in order to create or extend the services and products they offer. However, developing AI/ML systems has presented several engineering problems that are different from those that arise in, non-AI/ML software development. This study aims to investigate how software engineering (SE) has been applied in the development of AI/ML systems and identify challenges and practices that are applicable and determine whether they meet the needs of professionals. Also, we assessed whether these SE practices apply to different contexts, and in which areas they may be applicable. We conducted a systematic review of literature from 1990 to 2019 to (i) understand and summarize the current state of the art in this field and (ii) analyze its limitations and open challenges that will drive future research. Our results show these systems are developed on a lab context or a large company and followed a research-driven development process. The main challenges faced by professionals are in areas of testing, AI software quality, and data management. The contribution types of most of the proposed SE practices are guidelines, lessons learned, and tools.

## 1. Introduction

Recent technological advancements regarding cloud computing, big data management, algorithms, and tools have enabled a lot of opportunities for businesses, industries, and societies to make use of Artificial Intelligence. As such, Artificial Intelligence (AI), more specifically Machine Learning (ML) systems have widely been adopted by companies in all industries worldwide as value propositions in order to create or extend the services and products they offer (Rahman et al., 2019; Amershi et al., 2019). Almost all organizations today, from private to public sectors, are involved in some forms of AI initiatives (Bosch et al., 2020). The state-of-the-art AI/ML systems are quickly moving from a laboratory environment to an industrial one, and aim to increase the massive amount of accessible data available.

Due to the growing popularity of industrial AI/ML, a rapidly increasing amount of studies have been performed to understanding the processes, practices, and challenges faced by professionals in different companies (Byrne et al., 2017; Kim et al., 2018; Rahman et al., 2019; Amershi et al., 2019; Nascimento et al., 2019). Developing industrial AI/ML has come with several engineering problems that are different from those in traditional, non-AI/ML software development (Brynjolfsson and Mitchell, 2017; Byrne et al., 2017; Guo et al., 2019; Rahman et al., 2019; Amershi et al., 2019). For example, studies show some of the challenges that professionals face when developing AI/ML systems, for example: identifying customer business metrics, lack of a defined process, and data-centric engineering challenges (Brynjolfsson and Mitchell, 2017; Byrne et al., 2017; Nascimento et al., 2019).

In comparison to AI research, which focuses on algorithmic development or ML applications, engineering AI systems, i.e., tools, processes, and practices of managing, testing, and deploying ML models is a research area that has recently emerged (Kim et al., 2018; Amershi et al., 2019). In the last five years, we have observed that the area has rapidly evolved. A search on the Scopus database using the term (“artificial intelligence” OR “machine learning”) AND (“software development”) shows 2292 results, indicating a comparatively large amount of research articles about the topic.

Existing literature has revealed separate challenges faced by professionals in the development of ML systems and software engineering (SE) (Arpteg et al., 2018; Braiek and Khomh, 2020; Ghofrani et al., 2019; Kim et al., 2018; Lwakatare et al., 2019; Rahman et al., 2019; Amershi et al., 2019; Nascimento et al., 2019). Literature suggests that some practices are adopted by expert professionals or specialized software teams in AI/ML for different aspects of their engineering activities (Barash et al., 2019; Chakravarty, 2010; Ghofrani et al., 2019; Martin, 2016; Renggli et al., 2019; Amershi et al., 2019; Nascimento et al., 2019; Washizaki et al., 2019), however, they are often case-specific, and are due to a lack information regarding the context, and about problems on how these practices can be adopted. Hence, for both research and practice, it is necessary to have a consolidated body of knowledge that connects problems and SE practices that have been applied the development of AI/ML in industry.

In this paper, to cope with this lack of knowledge, we conducted a systematic literature review (SLR) on software engineering challenges, processes, practices, methods, and tools in the context of the development of AI/ML systems. We also aimed to identify the reported gap between current challenges and their solutions. Thus, we analyzed papers published between 1990 and 2019 with the aim of (i) understanding and summarizing the current state of the art in this field and (ii) analyzing the limitations and open challenges necessary to drive future research.

The overall question we were interested in was *“How is software engineering applied in the development of AI/ML systems?”*. From the objective, six research questions were derived as below:

RQ1. What types of AI/ML systems are investigated in the primary studies?
RQ2. What is the organizational context of the AI/ML system development?
RQ3. Which challenges are related to specific aspects of AI/ML development?
RQ4. How are SE practices adjusted to deal with specific AI/ML engineering challenges?
RQ5. What SE practices are reported in the primary studies?
RQ6. Which type of empirical methods are used?

The results we have identified could help other software organizations better understand which challenges are faced, and which SE practices may support the work of teams developing AI/ML systems. In this paper, our main contributions are as follows:

I. *The study provides a comprehensive view of the software industry regarding the development of AI/ML systems and lists the main challenges faced.*
II. *The study synthesizes the contextual setting of investigated SE practices, and infers the area of applicability of the empirical findings., It also presents the classification of SE practice types in the form of lessons learned, structure, guidelines, theory, tools, model, or advice. This can help to compare and generalize future research on SE practices for AI/ML systems.*

The remainder of the paper is divided into five sections. Section 2 introduces the background of the study. Section 3 presents how the research method is undertaken and the study execution. Section 4 reports the results and portrays our findings in relation to the research questions. In Section 5, we present the discussion regarding the results obtained and the challenges to the validity of the systematic literature review. Finally, in Section 6, we present the conclusions and intentions regarding future work.

## 2. Background

In this section, we list the studies focused on the practices and challenges faced in AI/ML system development processes and indicate which current literature reviews are available regarding SE for AI/ML.

### 2.1. *AI/ML software system*

A simple definition of Artificial Intelligence would be that it is the science of mimicking human mental faculties on a computer (Hopgood, 2005). Thus, AI is used to describe machines that mimic "cognitive" functions that humans associate with other human minds, such as "learning" and "problem-solving" (Russel and Norvig, 2012). AI systems include modules that present an opportunity to generate types of learning. ML is a type of artificial intelligence technique that makes decisions or predictions based on data, according to Agrawal et al. (2017). According to Bosch et al. (2020) the emerging field of AI engineering is an extension of software engineering, and one which includes new processes and technologies needed for the development and evolution of AI systems, i.e. systems that include AI components, such as ML. Thus, researchers have investigated the challenges, processes, and different ways to understand the processes involved in developing these systems.

### 2.2. *Challenges and Practices to AI/ML systems*

Previous research has examined software engineering challenges and practices for ML, such as Rahman et al. (2019), Amershi et al. (2019), Nascimento et al., (2019), Ghofrani et al. (2019), Braiek and Khomh (2020), Lwakatare et al. (2019), Arpteg et al. (2018), and Nguyen-Duc et al. (2020). However, they focused on presenting a set of challenges and practices analyzed separately in specific contexts. For example, Amershi et al. (2019) carried out a study with professionals working in different AI/ML teams in a large company, in this case, Microsoft. The authors' aim was to collect challenges and the best practices for SE that are used in the organization's internal projects. They reported different challenges and practices used in the lifecycle stages of the development of AI/ML projects. However, the set of challenges and practices reported by the authors is broad and the SE practices are basically a contribution of lessons learned in the projects, however, these practices are not actionable in other situations.

Sculley et al. (2015) investigated the risk factors of ML components in software projects and used the technical debt framework to explore these factors. The authors concluded that ML components are more likely to incur technical debts because they have the same maintenance problems as non-ML systems, as well as a set of problems specific to ML components. As a contribution, the authors present a set of anti-patterns and practices designed to avoid technical debt in systems that use ML components.

Rahman et al. (2019) performed a study in an industrial context and compiled a set of recommendations regarding SE practices found in the development of ML applications, the authors' summarized the practices in three distinct perspectives: (1) software engineering, (2) machine learning, and (3) industry-academia collaboration.

Munappy et al. (2019) present challenges applied in specific domains, such as data management for Deep Learning (DL) systems. The authors identified and categorized data management challenges faced by practitioners in different stages of end-to-end development. In addition, Arpteg et al. (2018) investigated SE challenges in DL systems and presented a set of twelve challenges categorized in three areas of development, production, and organizational challenges.

Nascimento et al. (2019) analyzed the challenges of ML system development in three small companies. The authors found that the challenges faced by the developers are, for instance, not having a defined process, difficulty identifying customer business metrics and difficulty designing the database structure. The authors suggest two checklists to support developers in overcoming the challenges faced in the stages of problem understanding and data handling.

Nguyen-Duc et al. (2020) proposed a research agenda for the continuous development of AI software. In their proposal, several challenges of maintaining the engineering practices and processes of AI software development at the continuous level are reported, i.e., AI software quality, requirement engineering, and tools and infrastructure.

However, none of the authors mentioned above explored the relationships and dependency between the challenges and practices of SE, or identify which contexts and which SE areas may be applicable, or even which challenges are met. Even though these studies provide an understanding of specific contexts of AI software development, they do not give an overall picture of AI challenges in different experimental or industrial contexts. Moreover, the SE practices recommended by the authors are not reproduceable, in other words, they are not actionable or can be used by professionals and in research. This motivates us to create a more useful systematic literature review.

### 2.3. *Existing literature reviews regarding SE for AI/ML*

In order to avoid repeating previous work, it is necessary to identify existing reviews on the topic. At the time of writing this study (summer 2020), we noted several literature reviews relevant to our topic. However, their objective, scope, and types of outcomes are not similar to what our intentions in this paper.

For instance, Zhang et al. (2020) conducted a review of 138 papers on testing in ML systems. The authors identified different ML test properties, test components, and test workflows. In addition, they summarized data sets used for experiments and the open-source testing tools/frameworks available, and analyzed the research trend, directions, opportunities and challenges in ML testing (Zhang et al., 2020).

Gonçalves et al. (2019) conducted an SLR to investigate intelligent user interface (IUI) design trends including interfaces generating by artificial intelligence systems, in the context of contemporary software systems, such as software systems based on the Internet of Things (IoT) or dedicated to smart cities. The preliminary results show that the models and technologies most used to develop IUIs are through tools, processes, frameworks. Interestingly, the most mentioned domain is health.

Washizaki et al. (2019) conducted an SLR and identified software quality methods and techniques for ML applications. The authors highlighted the problems in ML applications and revealed SE approaches and software testing research areas in order to address these problems. Nevertheless, there are systematic reviews on more specific domains of AI/ML systems, such as architectural and design standards for ML projects (Washizaki et al., 2019). The authors studied and classified patterns and anti-patterns of software architecture for ML, extracted from white and gray literature.

Hutchins et al. (2017) carried out a literature review on AI/Autonomous Systems (AS) design guidelines and a case study of the different incidences in which AI/AS systems can be involved in the production of many of the social and ethical dilemmas caused by the development of the system AI/AS.

However, ours is the first research that makes a comprehensive review of the literature on SE challenges and practices, and details the context in which these practices were applied and what type of contribution each practice proposes in order to support the development of AI/ML based on the classification of Paternoster et al. (2014). Furthermore, we analyze the challenges jointly with the proposed solutions.

## 3. Research methodology

We conducted an SLR to provide an overview of the state of the art on the topic of software engineering for supporting the development of AI/ML systems, following the guidelines from Kitchenham and Keele (2007), Kitchenham et al. (2009), and Kitchenham and Brereton (2013).

According to Wohlin (2014), systematic literature studies can include reviews and mappings and are a way of synthesizing evidence, for support researchers in understanding the actual status of a research area in software engineering. In our study, the result of the SLR is to obtain a broader view of the evidence and gaps in the SE methodologies, practices, and approaches that have been used in the context of the development of AI/ML systems.

In the next sections, the main stages of our process are presented according to the guideline from Petersen et al. (2015), including the search and study selection strategies, manual search, data extraction, quality assessment, and the data synthesis method.

### *3.1. Research goal and research questions*

The objectives of this SLR were defined as follows:

i. To investigate how software engineering has been applied in the development of AI/ML systems and identify challenges, approaches, practices, and methodologies that are applicable, and determine whether they meet the needs of professionals. Furthermore, we assessed whether these SE approaches and practices apply to different contexts, and which areas may be applicable.
ii. To synthesize scientific evidence on the use of methodologies, practices, and approaches when developing AI/ML systems.
iii. To identify the gap between current challenges and their solutions in order to be able to suggest areas for further investigation.

This SLR aimed to answer the following research question: "*How is software engineering applied in the development of AI/ML systems?*". To answer the main research question, the systematic literature review was divided into specific research questions, as described in Table 1.

**Table 1.** Research questions of the SLR.

| |
|---|
| RQ1. What types of AI/ML systems are investigated in the primary studies? |
| RQ2. What is the organizational context of the AI/ML system development? |
| RQ3. Which challenges are related to specific aspects of AI/ML development? |
| RQ4. How are SE practices adjusted to deal with specific AI/ML engineering challenges? |
| RQ5. What SE practices are reported in the primary studies? |
| RQ6. Which type of empirical methods are used? |

Particularly, this study presents an overview of which AI/ML domain applications are being developed (RQ1), for which context these applications are being developed (RQ2), what challenges are faced (RQ3), which SE practices (RQ4) are adapted and applied. In addition, for this RQ4, we verified in which areas of knowledge of SE (RQ5) the practices are being directed, based on SWEBOK (Bourque, P., Fairley, 2014). Finally, we verified which experimental methods are being applied (RQ6) in the primary studies, as shown in Fig. 1.

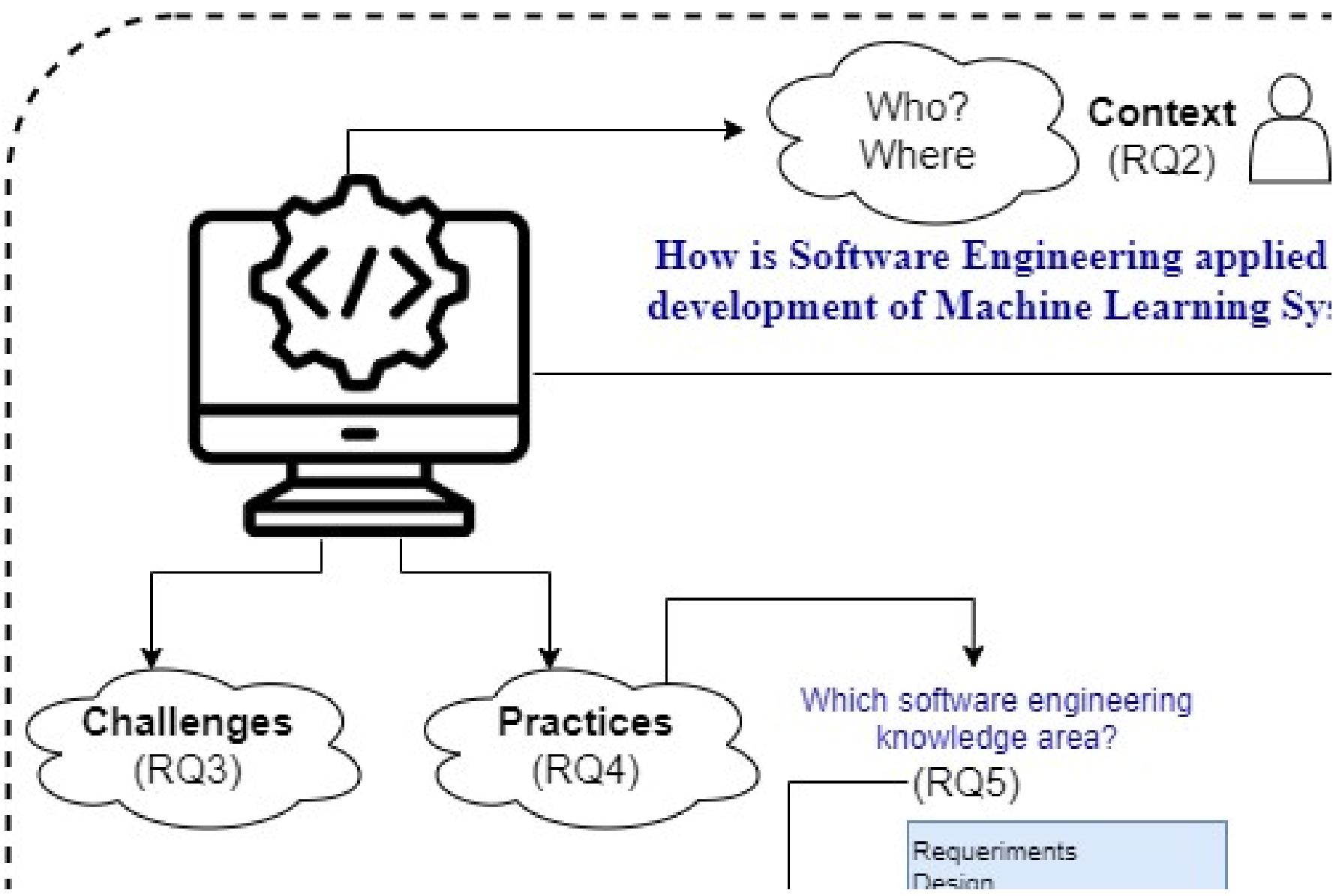

**Fig. 1.** Overview of the systematic literature review.

### 3.2. *Review procedure*

The review protocol defines the procedures used to conduct an SLR, which is important for the correct conduct and validity of the review/mapping (Wohlin et al., 2012). The systematic literature review was carried out in five stages, which are described below (Fig. 2):

**The first stage**: Free search and pilot search. First, we performed a free search to find papers about the topic SE for ML/AI. We read the full-text of the papers and selected control papers. After, we performed a pilot search in online databases, such as Scopus, IEEE, ACM, and Engineering Village to find an optimal search string and refined it. The searches helped to define the criteria for inclusion and exclusion and search string elaboration.

**The second stage**: Search strategy and study selection. Based on the search string, a total number of 2,057 unduplicated papers were retrieved. In this stage, based on the selection criteria three researchers classified a sample of 20 publications that were selected at random. Then, we evaluated the agreement between the researchers by applying, the Fleiss' Kappa statistical test (Fleiss, 1971). The result of this evaluation showed an almost perfect of agreement between the three researchers (kappa = 0.838) according to the range described by Muñoz and Bangdiwala (1997). In this way, it was possible to assess the degree of agreement between the researchers and the inclusion criteria for selecting papers.

**The third stage**: Additional manual search. The snowballing strategy, suggested by Wohlin (2014), was adopted to obtain the best possible coverage of the relevant literature for this research, and was used to find papers that presented challenges and SE practices applied in the development of AI/ML systems. This resulted in another 25 relevant papers.

**The fourth stage**: Data extraction and synthesis. From the primary papers, relevant data and information were transferred to an extraction form. A multi-step synthesis was performed to answer the main research questions. The following subsections detail the review protocol used to conduct this mapping.

**The fifth stage**: Quality assessment. To analyze the empirical evidence of the remaining papers, we performed a quality assessment (Table D.20).

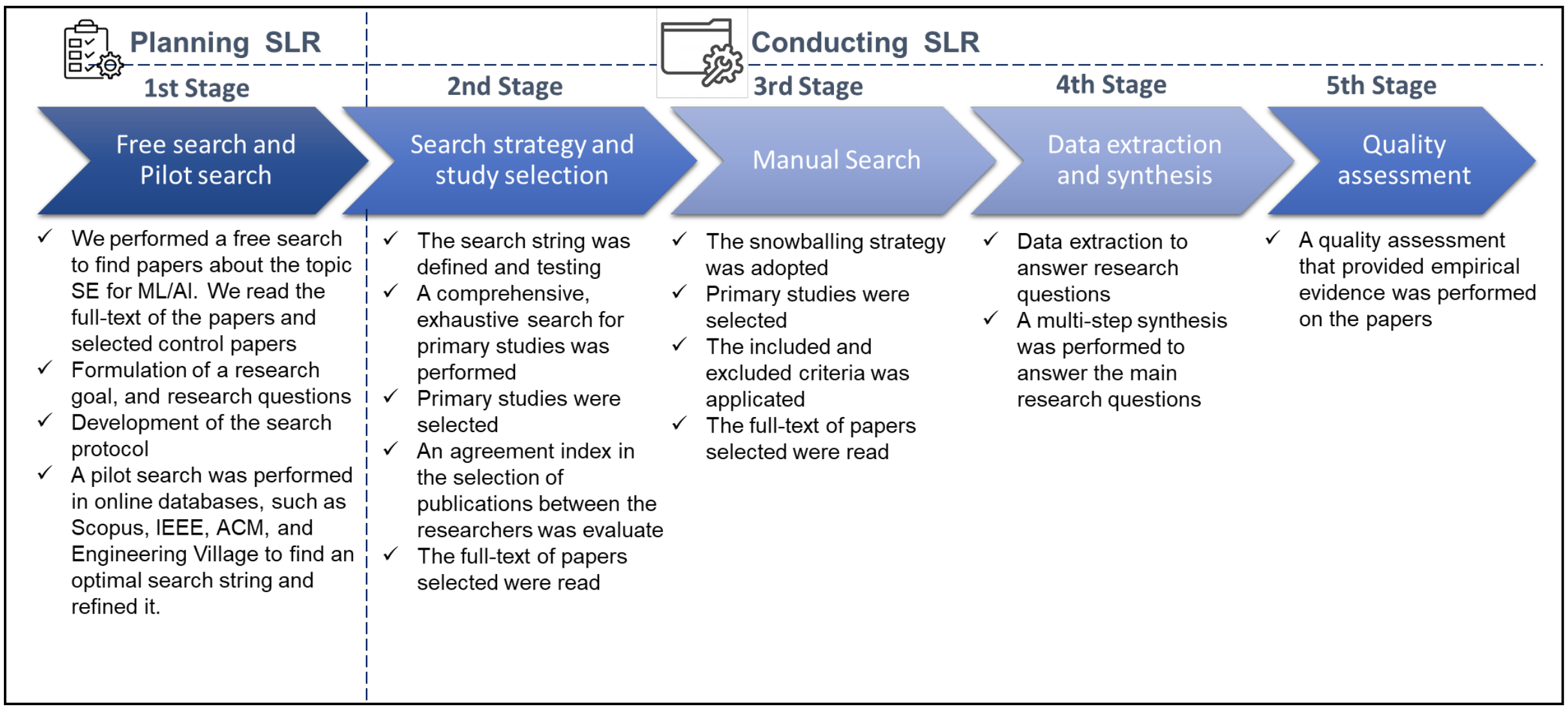

**Fig. 2.** Stages of systematic literature review, adapted from Kitchenham and Keele (2007) and Petersen et al. (2015).

### 3.3. *Data source selection and search strategy*

We conducted systematic literature review in the data sources Scopus, Engineering Village, IEEE, and ACM (Table 2). Scopus and Engineering Village libraries were included because they are meta-libraries and index publications from several well-known and reputable publishers, such as ACM, IEEE, Springer, and Elsevier. Even though the ACM and IEEE libraries were indexed by Scopus, these libraries were included to ensure that there were no publications excluded from Scopus' indexing of ACM and IEEE publications. ACM also indexes some publications from Springer Link and Science Direct. We selected English as search language, since those is the ones which the authors can understand.

**Table 2**. Searched data sources and number of retrievals (unduplicated).

| Data sources | Papers |
|---|---:|
| Scopus | 488 |
| IEEE | 593 |
| ACM | 968 |
| Engineering Village | 8 |
| Total | 2,057 |

### 3.4. *Search string*

When starting the research, we did some exploratory readings to become familiar with the type of information reported by the papers about the search topic. Our aim was to find keywords and guidance on inclusion and exclusion criteria, and refinement of the research questions.

To define the search terms, we used the procedure described by Kitchenham and Keele (2007), who suggest determining the parameters of population, intervention, comparison, outcome, and context, on the PICOC (Population, Intervention, Comparison, Outcome, Context), a methodology by Petticrew and H., Roberts (2006) described below: **Population** – *Machine Learning, Data Science, Data-Driven Application, AI application, and Artificial Intelligence*; **Intervention** – *Software Engineering, Software Requirement, Software Development, System Development, Software Application, Software Architecture, and Software Testing;* **Comparison** – Due to the objective of this research (mainly characterization), we did not apply any comparison parameters; **Outcome** – *empirical, case* *study, survey, questionnaire, interviews, action research, ethnography, Experience report, Experiment design*; **Context** – this does not apply, as there is no comparison to determine the context.

Therefore, we defined our search string as being strongly based on the **population, intervention,** and **outcome**. The first part of the search string (before the AND) has the purpose of limiting the results to the area of Machine Learning and Artificial Intelligence systems. The second part of the search string (after the AND) has the purpose of limiting the results to the areas of software engineering. Finally, the third part has the purpose of limiting the results to studies with empirical evidence.

The search string was constructed using the Boolean operators "OR" between the alternatives of writing and synonyms of the terms, and the Boolean operator "AND" to join the groups (Table 3). After the construction of the string, it was executed in all the data sources mentioned above (Table 2).

**Table 3.** Search string.

("Machine Learning" OR "Data Science" OR "Data-Driven Application" OR "AI application" OR "Artificial Intelligence") **AND** ("Software Engineering" OR "Software Requirement" OR "Software Development" OR "System Development" OR "Software Application" OR "Software Architecture" OR "Software Testing") **AND** ("empirical" OR "case study*" OR "survey" OR "questionnaire" OR "interview*" OR "action research" OR "ethnography" OR "Experience report" OR "Experiment design")

### 3.5. *Study selection*

The selection process for publications was composed of two stages, namely the 1st filter and 2nd filter. In the 1st filter, the researchers read only the title and the abstract to select publications related to software engineering applied in the development of AI/ML systems and applied the inclusion and exclusion criteria (Table 4).

**Table 4.** Inclusion and Exclusion Criteria.

| **IC** | **Description Inclusion Criteria** |
|---|---|
| **IC1.** | Publications that describe the use of methodologies, approaches, and practices software engineering in ML system development |
| **IC2.** | Publications that describe an AI software that includes an ML component to perform some intelligent functions and mentioned the use of SE methodologies, approaches, and practices |
| **EC** | Description Exclusion Criteria |
| **EC1.** | Publications that do not meet the inclusion criteria |
| **EC2.** | Publications in which the language is different from English |
| **EC3.** | Publications that are not available for reading or data collection (publications that are only accessible via payment or are not provided by the search engine) |
| **EC4.** | Duplicated publications |
| **EC5.** | Publications that focus on algorithmic aspects of AI/ML systems, but not engineering aspects |
| **EC6.** | Publications that theoretically discuss or propose practices or methods without empirical validation |
| **EC7.** | Publications about the use of AI/ML for improving software engineering practices |
| **EC8.** | Publications that are short papers and systematic review studies that did not perform an empirical study |

At this stage, by using the selection criteria, three researchers classified a sample of 20 publications selected at random. Then, the sample of 20 publications was analyzed by three researchers using the reading of title and abstract (1st filter). The purpose of analyzing this sample was to assess the agreement index in the selection of publications among the three researchers, using the Fleiss' Kappa statistical test (Fleiss, 1971). The result of this evaluation showed an almost perfect degree of agreement between the three researchers (kappa = 0.838), according to the scale described by Muñoz and Bangdiwala (1997).

In the 2nd filter, the researchers read the selected publications in full. The selection of publications in the 2nd filter was carried out using the same criteria used in the 1st filter. The study selection process is illustrated in Fig. 3 along with the number of papers at each stage. Fig. 3 shows the number of papers returned by each search library of the selected data source. In addition, it shows the total number of papers selected in the 1st filter and the total number of papers selected after the 2nd filter.

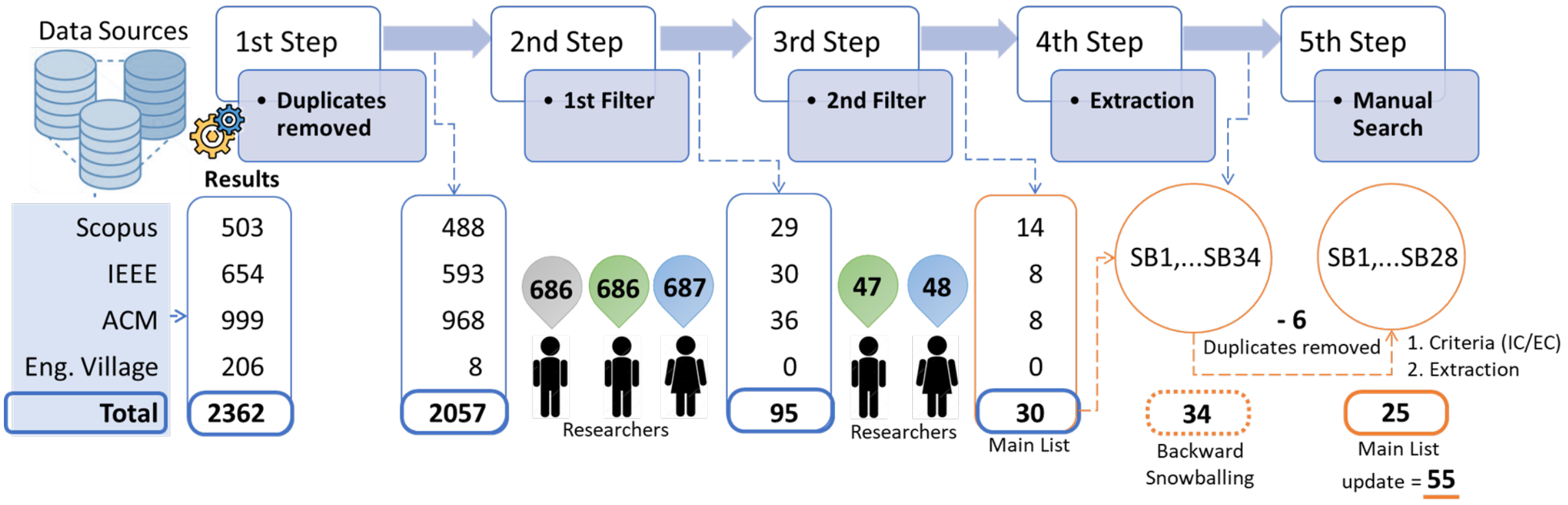

**Fig. 3.** Publication selection process.

The search string returned a total of 503 publications in the Scopus library, 654 in IEEE, 206 in Engineering Village, and 999 in ACM, and accounted for a total of 2,362 publications. Duplicate publications (305) were found during the process. In these cases, publications were counted only once. After removing duplicates, in total there were 2,057 publications. Thus, there were 488 publications in Scopus, 593 in IEEE, 8 in Engineering Village, and 968 in ACM.

During the first filter, in which the title and abstract of all publications are read, 1,962 publications were rejected, since they did not meet the inclusion criteria. The remaining 95 publications were read in their entirety and classified in the second filter, according to the same inclusion and exclusion criteria and the quality assessment. At the end of the process, 30 publications were accepted and extracted.

### 3.6. *Manual search*

The snowballing strategy was adopted to obtain the best possible coverage of the relevant literature for this research, and, as such, we looked for papers that presented challenges and SE practices applied in the development of AI/ML systems. The snowballing process, proposed by Wohlin (2014), presents two iteration processes: backwards and forwards snowballing.

At the end of the extraction of the 30 papers from the data sources, *backward snowballing* procedures were used, i.e., the reference list was used to identify new papers to be included (Wohlin, 2014). That is, for each paper extracted in the 2nd filter, the title of the paper cited and the context in which it was cited for inclusion in the snowballing list were verified. From the main list of papers, 34 papers were included in the snowballing list. Then, an analysis was carried out to check if there were duplicate papers, that is, with the same reference (title and authors) as the main list. Finally, 28 papers were included in the snowballing list for a complete reading. After reading the papers thoroughly and according to the same inclusion and exclusion criteria, 25 papers were accepted and extracted to the extraction formulary.

In total, 55 publications were included in the SLR, with 30 publications from data sources and 25 publications from snowballing as shown in Fig. 3. The selected papers were coded from P1 to P55 in Table A.17.

### 3.7. *Quality assessment*

Quality assessment of the primary papers providing empirical evidence was done after the final selection. In this SLR, we performed of the quality assessment process to assess the rigor, credibility, and relevance of the papers, which Kitchenham (2004) has identified as important for performing empirical research in software engineering.

Table 5 illustrates 10 quality evaluation criteria. For each criterion the papers met, they got a score of 1, and otherwise 0. If the paper satisfies the criteria to some extent, it gets a 0.5 point. This means that the maximum score a paper could get was 10. Thus, we can classify the paper's quality into the three categories from Berg et al. (2018) according to the total quality score as follows: low rigour (0 ~ 3.5), medium rigour (4 ~ 6.5), and high rigour (7 ~ 10). We selected all the papers and therefore, any study that scored 7 or below no was excluded from the list of selected studies. The complete table gives a summary of the scores obtained for each selected study can be found in Table D.20.

**Table 5.** Data quality assessment form.

| **Questions for assessing data quality from primary studies** |
|---|
| Screening questions<br>Q1 - Is the paper based on research (or is it merely a "lessons learned" report based on expert opinion)?<br>Q2 - Is there a clear statement of the aims of the research?<br>Q3 - Is there an adequate description of the context in which the research was carried out? |
| Research Design<br>Q4 - Was the research design appropriate in order to address the aims of the research?<br>Q5 - Was the population selected for the study appropriate to the aims of the research? |
| Data Collection & Analysis<br>Q6 - Was the data collected in a way that addressed the research issue?<br>Q7 - Was the data analysis sufficiently rigorous? |
| Findings<br>Q8 - Has the relationship between researcher and participants been considered to an adequate degree?<br>Q9 - Is there a clear statement of findings?<br>Q10 - Is the study of value for research or practice? |

### 3.8. *Data extraction and synthesis*

The papers that were selected for the 2nd filter followed the strategy for data extraction. In the 2nd filter, the data extraction process was done by reading each of the publications selected in the 1st filter. To start this process, an extraction form was defined to record the necessary information related to each publication, reducing the opportunity to include researcher bias. Note that this data extraction form is a result of a keywording process. According to Fernandez et al. (2011), this strategy ensures that the same data extraction criteria are used, facilitating their classification. Thus, during data extraction, the researcher recorded a set of possible answers that should answer the research questions. Thus, the information was extracted according to each research question (Table 6).

**Table 6.** Data extraction form.

| Data item | Value | RQs |
|---|---|---|
| **Paper ID** | Paper identifier (Number) | |
| **Paper title** | Name of the paper | |
| **Paper authors and affiliations** | Paper authors | |
| **Venue** | Name of publication venue | |
| **Venue type** | Type of publication venue | |
| **Country** | Name of country | |
| **Year of publication** | The year the paper was published (Calendar year) | |
| **Research strategy 1** | Which was the research strategy used according to Stol and Fitzgerald (2018) | RQ6 |
| **Research strategy 2** | Which was the research strategy used according to Storey et al. (2020) | RQ6 |
| **Type of ML/AI systems** | Types of ML/AI systems investigated in the primary studies | RQ1 |
| **Organizational context** | Details about the organizational context of the ML/AI system development | RQ2 |
| **Challenges** | Challenges in the paper related specific aspects of ML/AI development | RQ3 |
| **SE Practices** | SE practices mentioned in the primary studies | RQ4 |
| **SWEBOK** | Knowledge area mentioned in the primary studies for SE practices | RQ5 |
| **Research method** | Type of empirical methods used (case study, survey, experiment, etc) | RQ6 |
| **Quality assessment** | Quality items assessed in the paper | |
| **Interesting parts** | Interesting parts discussed in the paper | |
| **References** | For snowballing | |

### 3.9. *Keywording processes*

The goal of keywording is to efficiently create a classification scheme, ensuring that all relevant papers receive consideration and provide input to the data extraction form. A good classification scheme should base on established definitions, normally found in the literature. The scheme should be orthogonal, which means it should be composed of clear boundary elements. Furthermore, they should be complete and widely accepted by the research community. Our keywording process is based on a set of pilot papers. We progressively repeat the process for each paper until reaching saturation.

- Step 1: Reading the abstracts, keywords, and full-text of the selected papers and assigning them a set of keywords to answer the asking questions. This step is similar to the open coding of grounded theory (Charmaz and Bryant, 2010).
- Step 2: Organizing the keywords for each RQ, merging close keywords, adding missing keywords from literature to make the list of keywords complete for each RQ.

As a result of the keywording process, we come up with the list for RQ1 and RQ2, as shown in Table 7 and Table 8.

**Table 7.** Keywording for RQ1.

| **Category** | **Keywording** |
|---|---|
| *Application domain*: aims to identify the application domain that the developed system must meet | Finance, health, agriculture, education, software, logistics, society, living, others, and unknown |
| *Learning paradigm*: classifies the learning paradigm (s) mentioned in the development of ML systems in organizations. | Supervised learning, unsupervised learning, reinforcement learning, multi-task learning, others, unknown |
| *Machine learning approaches*: classifies the ML approach (es) mentioned in systems development | Classification and regression trees, neural network, probabilistic graphical networks, kernel methods, others, unknown |

For **RQ2,** we grouped the organizational context data into four categories:

**Table 8.** Keywording for RQ2.

| **Category** | **Keywording** |
|---|---|
| *Organization type*: indicates the size of the organization that the AI/ML system was developed for | Large companies, small companies/start-ups or laboratories and unknown |
| *Development process*: indicates the development process used by the organization to develop the AI/ML system. | Research, ad hoc, waterfall, and agile |
| *Organizational context*: indicates whether the system is developed by an internal team from the organization or by a third party | In-house, outsource |
| *Data source*: indicates the source of the data that was used for the development of the system | Open source, private or experimental/simulated data |

Regarding RQ5, there is no need for a keywording process, since the knowledge areas are extracted directly from SWEBOK (Bourque, P., Fairley, 2014), and the practices, or types of contributions, followed the classification of Berg et al. (2018). RQ3 and RQ4 are open questions and answers for them were extracted according to a proper open coding process (Cruzes and Dybå, 2011). RQ6 uses pre-defined research categories used in existing SE studies to according to Stol and Fitzgerald (2018) and Storey et al. (2020).

### 3.10. *Thematic analysis*

To answer RQ3, we performed a thematic analysis of the citations referenced in the papers on the challenges faced in the development of AI/ML according to Cruzes and Dybå (2011). First, we extracted the relevant text reporting challenges when developing AI/ML, based on the information included in the extraction form. Then, for each challenge, we create codes according to the content analysis of the reported challenges. Finally, the entire coding process was reviewed by two other researchers, to evaluate the results obtained in the data analysis. After reviewing the codes, we used the Nvivo[1] tool to organize the encoded data. Based on the codes created, categories and subcategories were generated by Nvivo.

Thus, we organized the challenges into categories and subcategories that emerged from the data. At the end of this process, 13 categories were generated by encoding the data using Nvivo. Each category unites subcategories that are related to the main (parent) category.

Table B.18 summarizes all the categories and subcategories that emerged from the data through the coding performed by Nvivo. It also includes the total number of papers that references the challenges (category or subcategory), the identification of the paper that references the challenge, and the number of references of the challenge mentioned in the paper. Each category was generated from the references most evident in the papers. In addition, each category unites subcategories that are related to the parent category. The main categories group the challenges related to a theme described in Subsection 4.4.

## 4. Results

This section is divided into two subsections. Subsection 4.1 presents the publication frequency of primary studies from 1990 to 2019. Subsections 4.2 to 4.7 presents the extracted data of the primary studies. The subsection is organized into six research questions to allow for better visualization and presentation of the most relevant findings.

### *4.1. Demographics of primary studies*

Demographics of the selected primary studies are summarized in Fig. 4, Fig. 5, and Table 9. Fig. 4 shows the number of studies published between 1990–2019 in relation to software engineering for AI/ML and constitutes a total of 55 published papers.

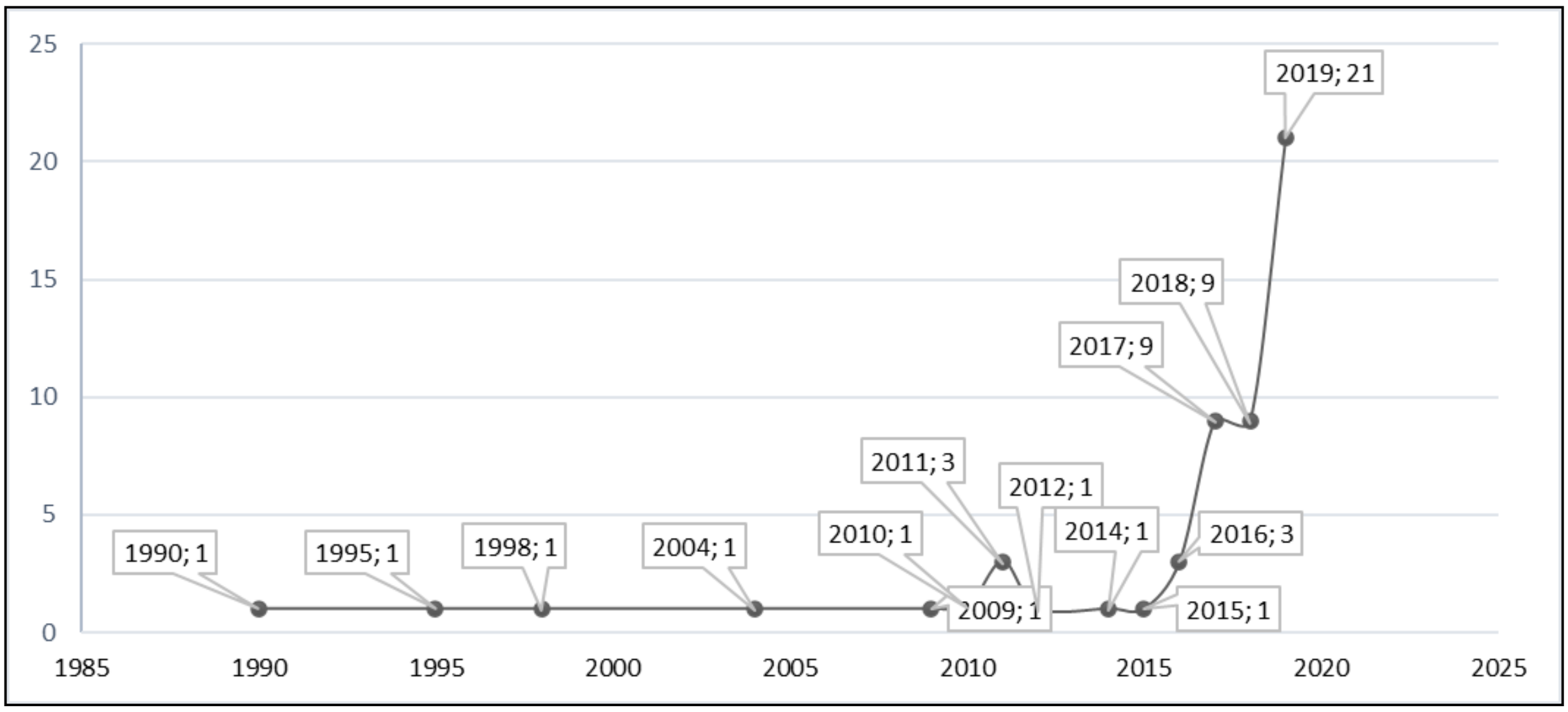

**Fig. 4.** Publication frequency, 1990–2019.

Observing the distribution of studies from a temporal perspective, there were no extensive studies in the area of software engineering for AI or ML until 2016. However, in 2011, 2012, and 2016, there were at least two or three studies carried out in these years. In 2017 and 2018, there was a remarkable growth curve, with nine publications in this area. But, in 2019, there was a peak of high growth of publications and 21 studies published this year. This result shows that in the last three years (2017, 2018, and 2019) the SE community has realized the need for research focused on AI and ML systems. This SLR was carried out in February 2020, so for the current year it found only one study.

[1] Nvivo - https://www.qsrinternational.com/nvivo-qualitative-data-analysis-software/home

We assessed the distribution of published studies using (a) type of publication (journal, conference, or gray literature) and (b) targeted publication locations (which are the most published journals and conferences). Fig. 5 shows the temporal distribution according to the type of publication (a). The most common type of publication is conference papers with 43 (79%) studies, followed by journal papers with 7 (13%) and, finally, gray literature with 5 (9%) studies. Such a high number of papers at conferences may indicate that this topic is relevant, although some studies have already been published in the years before 2017, however to a lesser degree.

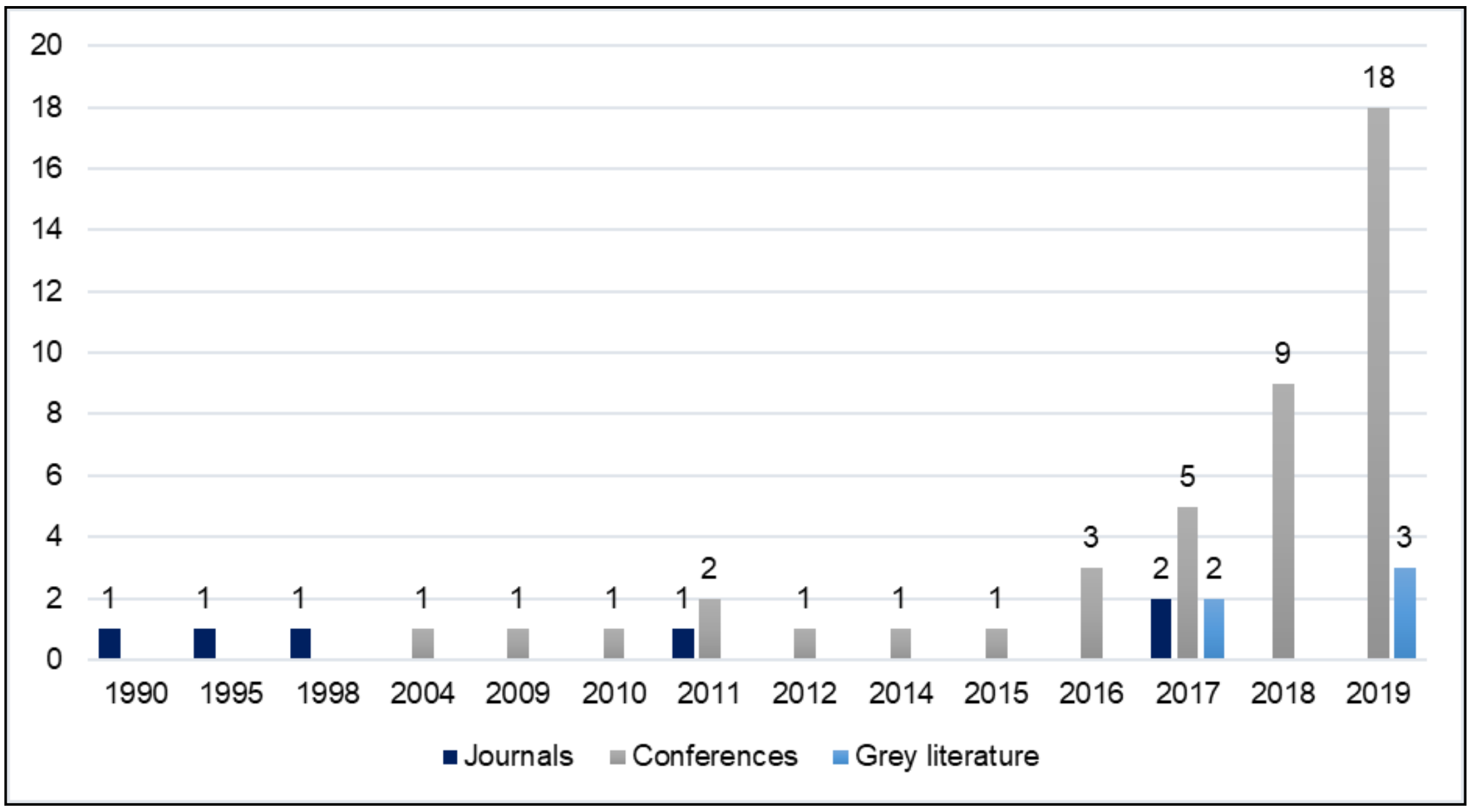

**Fig. 5.** Publication frequency by type.

Table 9 **Error! Reference source not found.**shows the origin of the publications by type. We saw that the research is spread over 34 different venues (55 papers). This result is an indication that SE for AI and ML is perceived as emerging research in many areas of research, rather than a specific research topic.

**Table 9.** Publication venues.

| **Publication venues** | **Type** | **Reference** | **Papers** |
|---|---|---|---|
| International Conference on Software Engineering (ICSE) | Conference | P2, P9, P18, P24, P39, P55 | 6 |
| International Conference on Automated Software Engineering (ASE) | Conference | P15, P44, P48, P53 | 4 |
| European Software Engineering Conference and Symposium on the Foundations of Software Engineering (ESEC/FSE) | Conference | P20, P21, P22, P51 | 4 |
| Euromicro Conference on Software Engineering and Advanced Applications | Conference | P8, P30, P32 | 3 |

| Publication venues | Type | Reference | Papers |
|---|---|---|---|
| (SEAA) | | | |
| Conference of the Computers in Education Society of Ireland (CESI) | Conference | P3, P26 | 2 |
| International Symposium on Empirical Software Engineering and Measurement (ESEM) | Conference | P1, P27 | 2 |
| Systems and Software | Journal | P36, P52 | 2 |
| Others (venues with 1 publication) | - | P4, P5, P6, P7, P10, P11, P12, P13, P14, P16, P17, P19, P23, P25, P28, P29, P33, P34, P35, P38, P40, P41, P42, P43, P47, P49, P50 | 27 |
| Grey literature | - | P31, P37, P45, P46, P54 | 5 |
| | | Total: | **55** |

In Fig. 6 the x-axis represents the publication venues, while the y- axis represents the rigour. Only four papers obtained low rigour score (7%), as they did not provide enough details about the research design, sampling, data collection, data analysis, and assessment of the validity of the results. However, in general, the papers obtained high rigour score (76%), indicating that the quality of the research was high. The figure is based on the quality assessment presented in Table D.20.

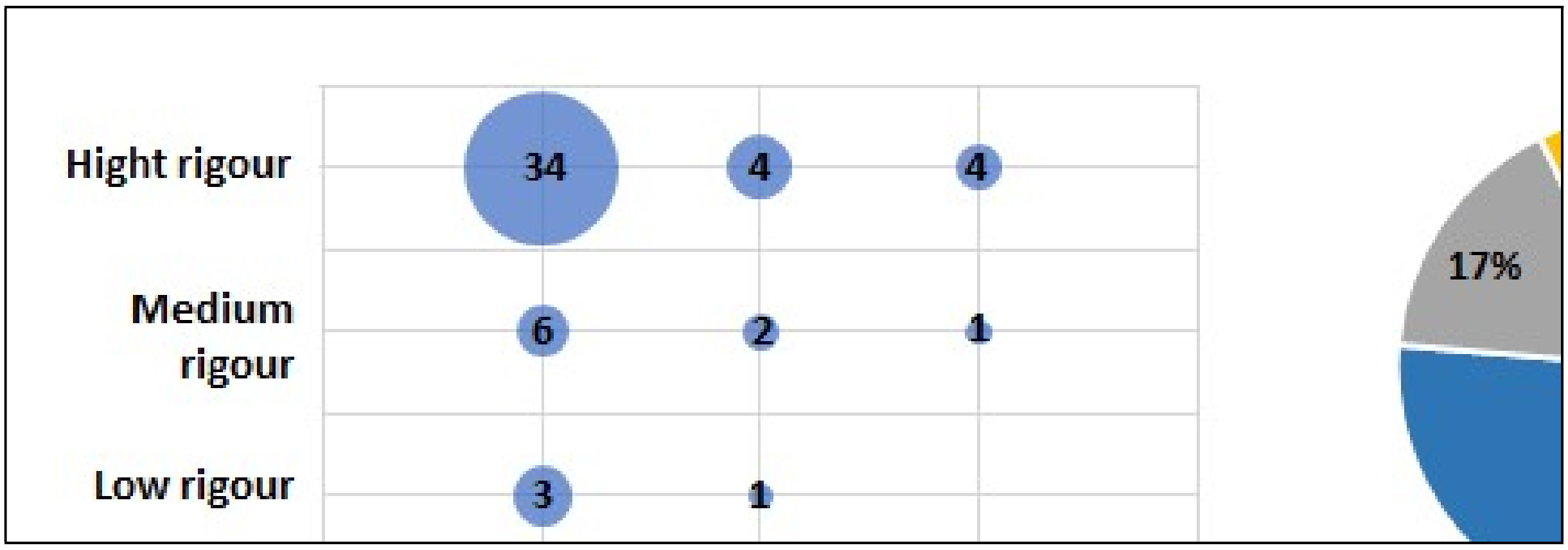

**Fig. 6.** Rigour of each publication venues and paper score.

*4.2. RQ1. What types of AI/ML systems are investigated in the primary studies?*

The answers to RQ1 are given in Table 10, Fig. 7, and Fig. 8. In the first category (i) Application domain, Table 10 **Error! Reference source not found.**shows the domains mentioned in the studies with the identification of the papers and the total of each domain. Only 40% of the total of primary studies showed the mastery of the application of the AI/ML systems evaluated or developed during the research. In other words, a large amount of empirical research about AI development does not provide sufficient contextual information about the application domains they are dealing with.

**Table 10.** Application domain for AI/ML systems.

| Automotive | Finance | Healthcare | Software | Education | Others | NA |
|---|---|---|---|---|---|---|
| P3, P5, P7, P8, P43, P48, P55 | P1 (2), P3, P5 | P3, P4 (3) | P2, P3, P5 | P3 | P3, P51, P45 | P10, P12, P15, P16, P21, P23, P24, P25, P26, P29, P30 |
| **7** | **4** | **4** | **3** | **1** | **3** | **11** |

The popular application domains that we observed included automotive (7 papers), finance (4 papers), and healthcare (3 papers). In the second category, (ii) *Machine Learning paradigms*, Fig. 7 presents the most applied paradigms in the development of AI/ML systems. Among the paradigms used, the first is supervised learning with 66% of cases. Second, comes reinforcement learning with 12% of cases. Third, unsupervised learning with 9% of cases.

- Example of supervised learning: prediction of house price, weather, and air quality forecast
- Example of unsupervised learning: finding customer segments out of marketing data, representative feature selections for a complex dataset

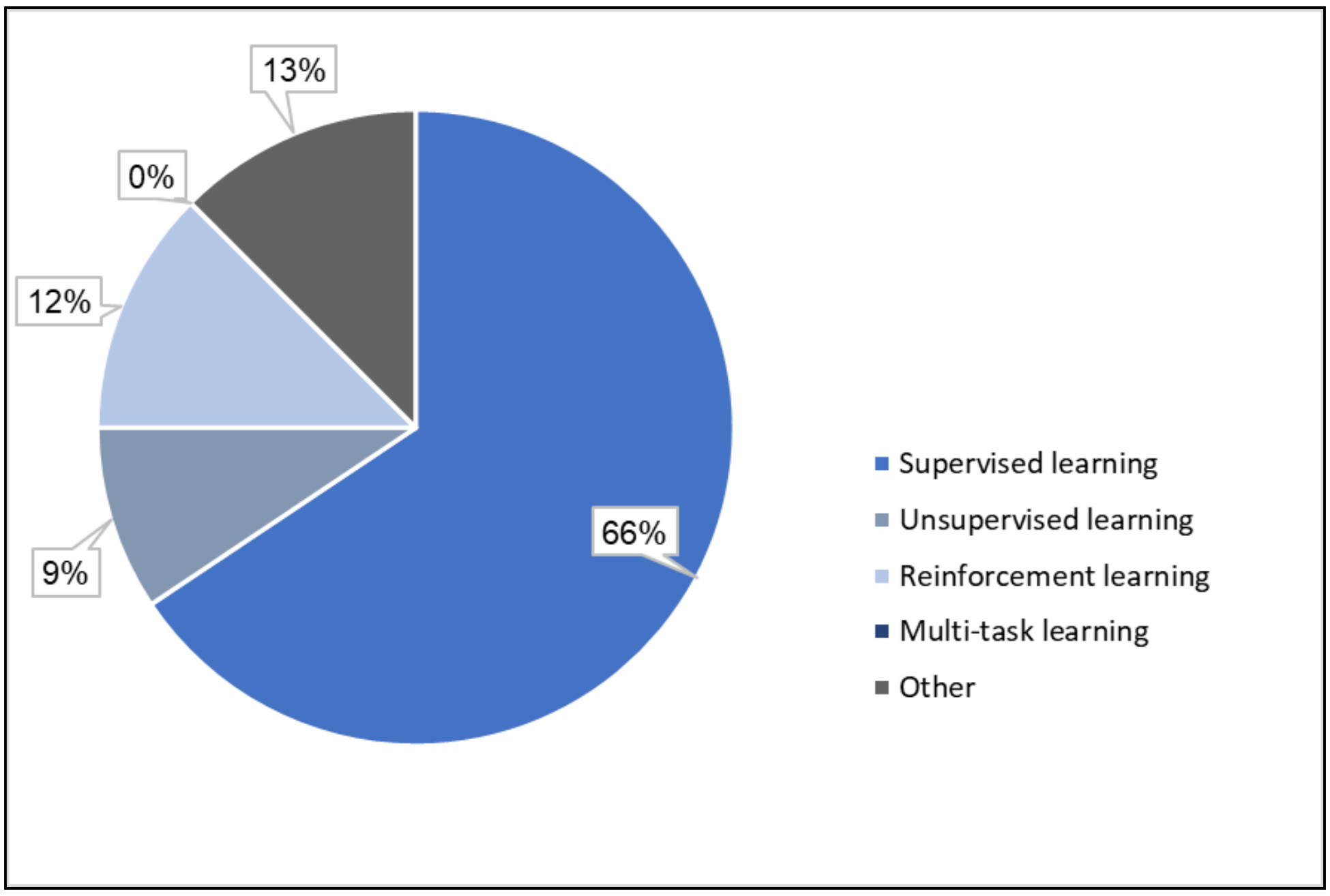

**Fig. 7.** Machine Learning paradigms.

In the third category, (iii) *Machine Learning approaches*, Fig. 8 presents the most used approaches in the development of these systems. We understand that there might be a difference in terms of speed, interpretability, and accuracy among the approaches, and this depends on the context and available competence with which a suitable approach is selected. Among the surveyed approaches, the most cited are Artificial Neural Networks (ANNs)/Deep Learning (DL) with 40% of cases, classification and regression with 31% of cases, and probabilistic models with 6% of cases. It is important to highlight that in the two categories mentioned, many cases do not mention which paradigm (12%) and approach (23%) were used.

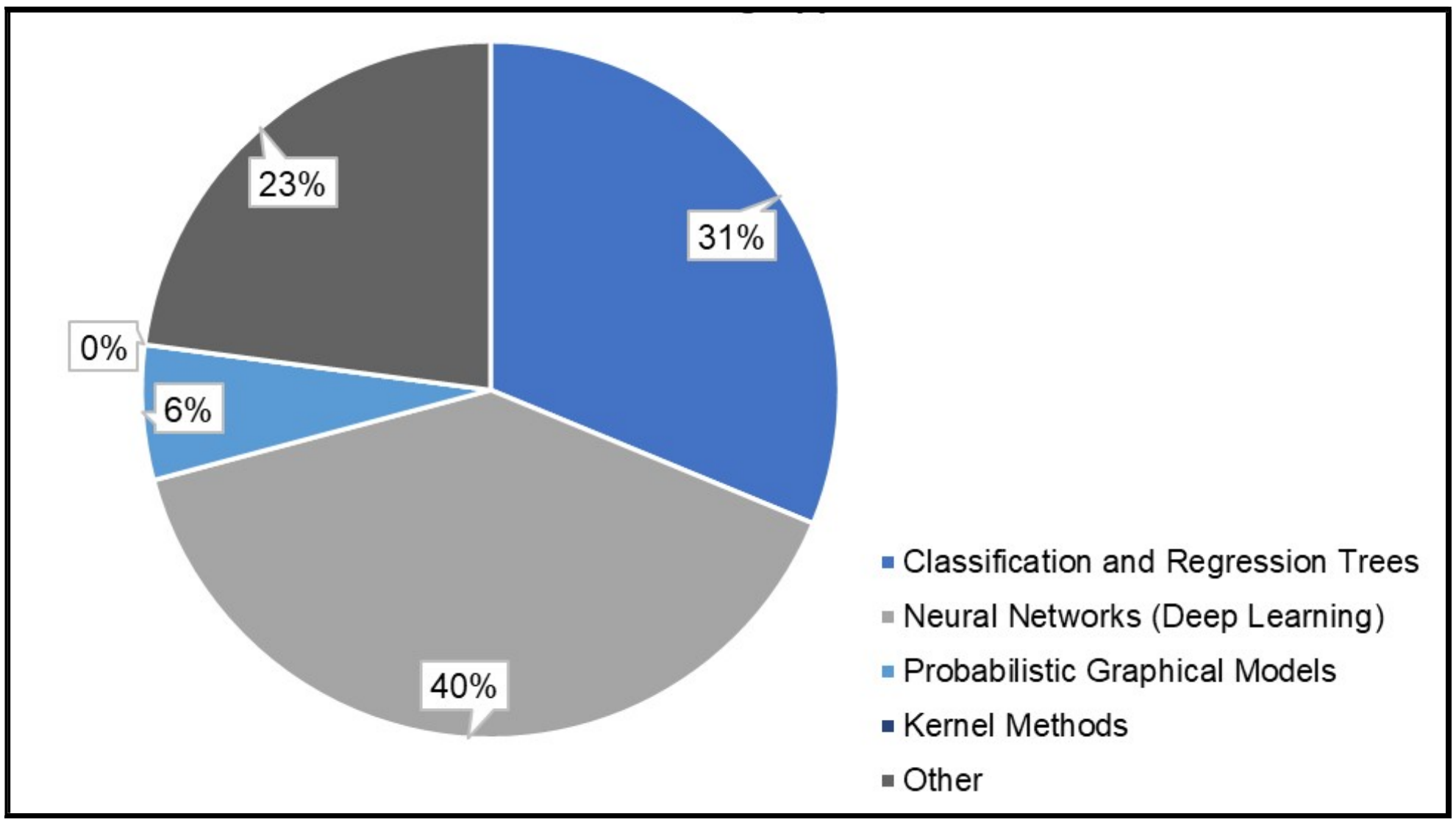

**Fig. 8.** Machine Learning approaches.

**Take away 1: Empirical studies report AI software in various application domains, with the main focus being Automotive, Finance, and Healthcare and with the wide adoption of Artificial Neural Network algorithms**

*4.3. RQ2. What is the organizational context of the AI/ML system development?*

Addressing RQ2, Fig. 9 shows the distribution of data by the company's organizational context. Most AI/ML systems are developed in the laboratory (26 cases - 41%) and large companies (21 cases - 33%). Also, there are 5 cases (8%) of systems developed by small businesses, and 11 cases (16%) were not applicable/did not report their development process (NA) in the papers.

As for the development process used by the organization to develop the AI/ML system, 38 (58%) primary studies reported AI/ML development as research processes, 10 (15%) studies showed them as informal professional processes (ad hoc), one case reported Waterfall AI development, and two studies discussed Agile methodology. Finally, 10 (15%) studies did not report (NA) their development process, and five studies have other development processes.

As for the organizational context in which the system is developed, whether by an internal or external team, in the majority of primary studies (38 cases – 60%), the systems are developed in-house by the AI engineers and software engineers of the organization. Only two cases showed that there were outsourced teams that developed the AI/ML systems. In 23 (37%) primary studies, no organizational context was reported.

Finally, regarding the data source used for the development of these systems, most organizations use their own private data (21 cases – 37%). However, open-source data (11 cases -19%) and experimental data (9 cases – 16%) are also used. Finally, 16 (28%) studies did not report the data source used.

**Take away 2: The majority of our research on SE reports AI development as a research-driven process. Significant AI projects are reported in a lab context or a large company. There is a lack of process studies in small and medium enterprises (SME) or start-up contexts.**

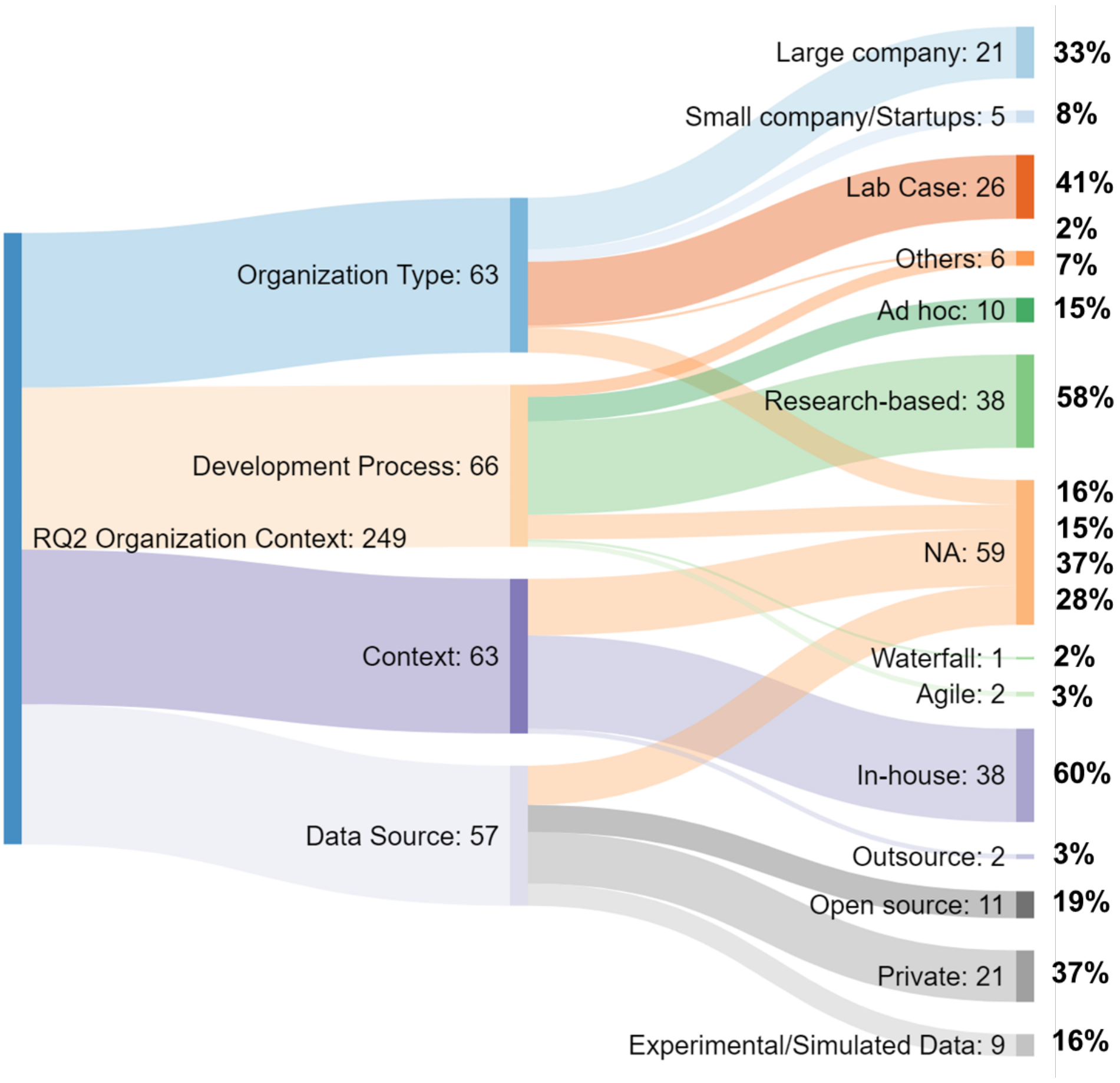

**Fig. 9.** Organizational contexts for the development of AI/ML Systems.

### *4.4. RQ3. Which challenges are related to specific aspects of AI/ML development?*

Fig. 10 presents a hierarchy graph that shows the number of references from different codes with the categories most referenced in the data. The more coding a category has, the larger its area. In addition, the subcategories (the child codes) are grouped into the parent category. Below, we present the categories of challenges ordered by their popularity, including Software Testing (30 references) and AI Software Quality (27). Followed by the categories of Model Development (16), Data Management (16), Project Management (15), Infrastructure (14), and Requirements Engineering (13). The categories of 10 to 6 references were AI Engineering (10), Architecture Design (8), and Model Implementation (6). The categories that had up to two references were Integration (2), Operational Support (1), and Education (1). Table B.18 presents the challenges faced by professionals in the development of AI/ML systems, and in it we present the most evident categories and subcategories, with the highest number of citations.

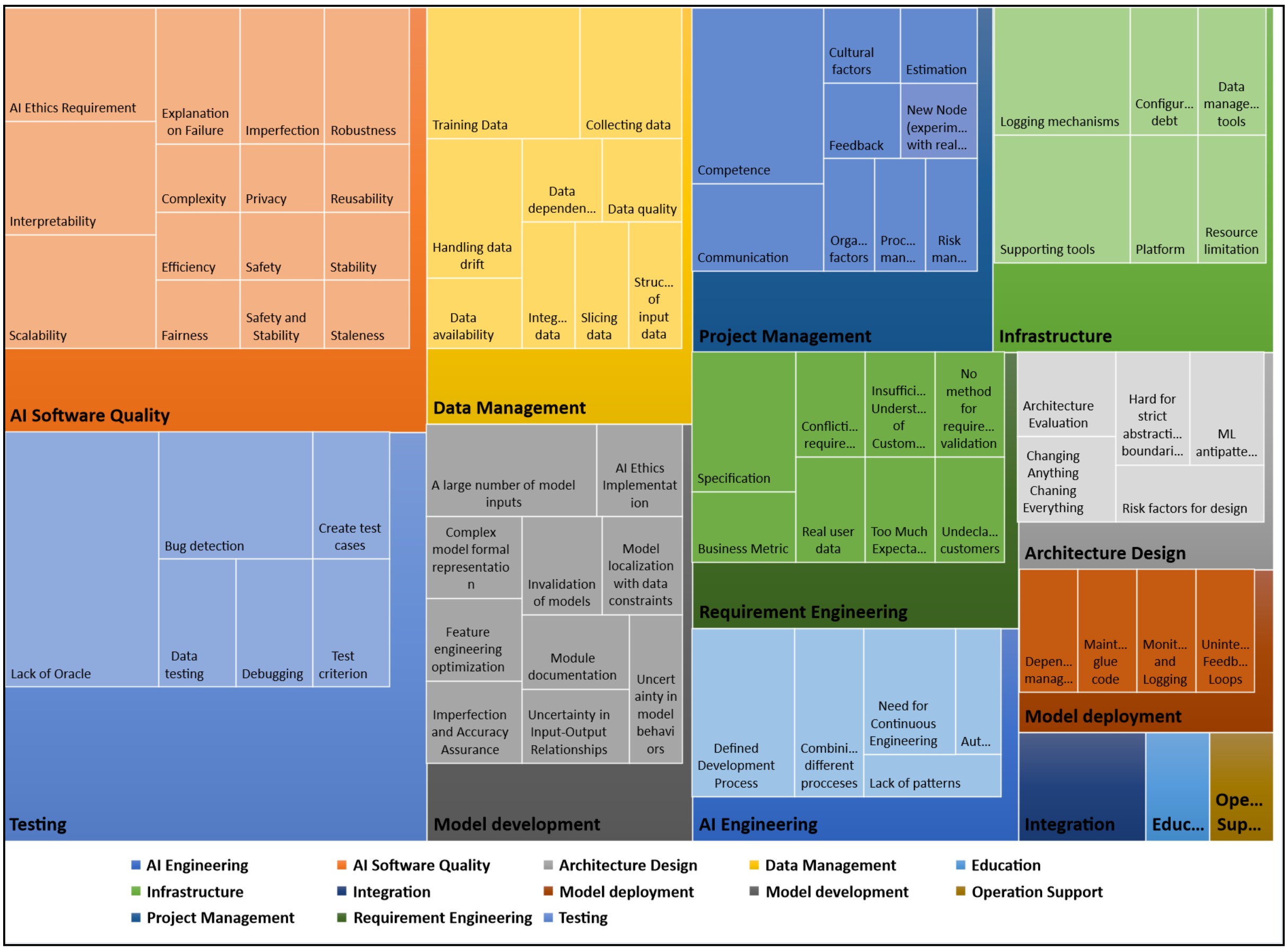

**Fig. 10.** Overview of the main challenges faced in the development of AI/ML systems.

*4.4.1. Software Testing*

Testing includes the verification of the AI software or ML model to see if it behaves as expected. The need to validate behaviour of modules using ML, Deep Learning, Recurrent Neural Network, etc. is acknowledged in many primary studies. It is not surprising that the "Software Quality" and "Tests" categories are the most cited topics as challenging in the development of ML/AI systems. Alshangiti et al. (2019), Guo et al., (2019), Rahman et al., (2019), and Amershi et al. (2019) also reported that performing tests on ML systems is challenging because the procedures to perform the tests on these systems are different from the traditional techniques known and applied in non-ML systems. Traditional testing techniques and methods that aim to increase structural coverage towards the exploration of diverse software states is not very useful for DL systems (Kim et al., 2019; Rahman et al., 2019; Xie et al., 2011). The current understanding about characteristics of bugs found in ML systems is currently limited (Thung et al., 2012), and leads to the need for systematic and efficient bug detection and fixing approaches.

Challenges related to testing regarding different aspects of AI/ML include bug detection, creating test cases, data testing, debugging, and test criteria. In AI/ML systems, the rules are inferred from training data (that is, they are generated inductively) and not deductively as occurs in non-ML systems (Braiek and Khomh, 2020; Ishikawa, 2019; L. Ma et al., 2018; Xie et al., 2011). "Bug detection" was also one of the most cited main subcategories (Braiek and Khomh, 2020; Selsam et al., 2017). A bug in an ML program may come from its training data, program code, execution environment, or third-party frameworks (Braiek and Khomh, 2020). Compared with traditional software, the dimension and potential testing space of an ML program are both much larger. This leads to the problem of estimating and assigning budgets and efforts to AI/ML testing activities.

Implementation errors are difficult to detect and resolve, as there can be many other potential causes of unwanted behaviour in an ML system. For example, an implementation error can lead to incorrect gradients and cause a learning algorithm to be interrupted, but this symptom can also be caused by noise in the training data, poor model choice, unfavourable optimization scenario, research strategy inadequate, or numerical instability (Selsam et al., 2017).

Another significant challenge in testing AI software is the lack of oracle (Braiek and Khomh, 2020; Ishikawa, 2019; L. Ma et al., 2018; Xie et al., 2011). It is difficult to clearly define the correctness criteria for system outputs or the right outputs for each individual input.

**Take away 3: Testing space for AI software is much larger, more heterogeneous and, in many cases, it is difficult to formally define in comparison to traditional software testing.**

*4.4.2. AI Software Quality*

In the challenge of obtaining "AI Software Quality" in AI/ML systems, three main subcategories stand out: (i) defining the ethics requirements in AI, (ii) interpretability, and (iii) scalability are perceived as the most challenging aspects in these systems (Table 11). For AI ethics requirements, the methods, and tools to implement AI ethics are lacking (Bryson and Winfield, 2017; Vakkuri et al., 2019; Vakkuri and Kemell, 2019). Furthermore, ethics is not a formal requirement in AI projects and developers are not well informed or aware of how ethics should be applied in AI projects, so they use known practical methods (Vakkuri et al., 2019; Vakkuri and Kemell, 2019). Regarding interpretability, the activity of labelling data is related to the transparency of the ML model (Arpteg et al., 2018; Braiek and Khomh, 2020; Lwakatare et al., 2019). In the activity of labelling data, there is a need for processes and tools to accurately and consistently label large data sets (Lwakatare et al., 2019). Due to a lack of transparency in the ML model, the model is inherently irreducible, so an accurate explanation of the model will be as complex as the model (Arpteg et al., 2018). Furthermore, when it comes to implementing ML algorithms, ML engineers sometimes struggle to understand these formulas, rules, or concepts, and often require sophisticated numerical calculations that can be difficult to implement in sophisticated, high performance hardware architectures (Braiek and Khomh, 2020). As for scalability, the challenges are in the process of integrating ML models into scalable software systems (Amershi et al., 2019) and data quality assessment (DIng et al., 2017; Kim et al., 2018). One of the reasons for obtaining poor data quality is due to flaws in the data collection and sampling procedure.

**Table 11.** Main quality attributes of AI.

| **Quality attributes** | **Description** | **Challenges** | **Mentioned in** |
|---|---|---|---|
| Interpretability | The goal of interpretability is to describe the internals of a system in a way that is understandable to humans (Gilpin et al., 2019) | Lack of model transparency, the model is inherently irreducible<br>Lack of supporting tools and methods | Lwakatare et al. (2019), Arpteg et al. (2018), Braiek and Khomh (2020), Ishikawa (2019), S. Ma et al. (2018) |
| AI Ethics Requirement | Transparency, accountability, and responsibility, as well as | Ethical concerns are neglected by developers | Vakkuri et al. (2019), Vakkuri and Kemell |

| Quality attributes | Description | Challenges | Mentioned in |
|---|---|---|---|
| | predictability as a subset of transparency, are the main principles of AI ethics (Vakkuri et al., 2020) | Lack of a formal requirement process to design ethics in AI projects. Lack of approaches, techniques, and tools for tackling ethics requirement | (2019), Bryson and Winfield (2017) |
| Scalability | The size of input data, the amount of variables | Batch processing is time-consuming Iterative workflow can be expensive Quick visualization of large data sets is complicated | Amershi et al. (2019), DIng et al. (2017), Kim et al. (2018) |
| Imperfection | Limited input information | It is very difficult to generate adequate outputs for any of various inputs (i.e., 100% accuracy | Ishikawa (2019), Tamir and Kandel (1995) |
| Safety | Safety in terms of avoiding harm from unwanted outcomes | Implementing empirical risk minimization in the AI model is uncertain | Ishikawa (2019), Varshney (2017) |
| Robustness | The effectiveness of algorithms while being tested on new independent datasets | Lack of robustness in DL models | Guo et al. (2019), Sun et al. (2018b) |
| Complexity | Number of adjustable features (dimensionality of space) | Effort required for specifying and building models in some domains is high due to the complexity | Jana et al. (2018) |
| Efficiency | Ability to produce outcomes with a minimum amount of waste | How to reduce the time cost memory consumption after model migration and quantization | Guo et al. (2019) |
| Fairness | (1) anti-classification, (2) classification parity and (3) calibration (Corbett-Davies, Sam, 2018) | Dealing with individual discrimination in given ML models | Aggarwal et al. (2019) |
| Stability | How ML is perturbed by small changes to its inputs | Difficulty in maintaining tasks due to ML's behaviour can be determined by both program codes and input data | Yokoyama (2019) |

| Quality attributes | Description | Challenges | Mentioned in |
|---|---|---|---|
| Staleness | Predictive power of an ML model decreasing over time | Dealing with staleness when operating models with changing data trends | Yokoyama (2019) |

**Take away 4: Besides Interpretability, empirical research highlights various engineering challenges in regard to AI ethics.**

*4.4.3. Data management*

State-of-the-art AI/ML systems rely on high-effort data management tasks, such as data exploration, data preparation, and data cleaning. Challenges regarding the data collection, processing, data availability, and quality are highlighted in our primary studies. The lack of data, the lack of values, the delay in sending data, the lack of metadata, the granularity of data, the scarcity of different samples are challenges related to the availability of data for ML projects. Other challenges are data manipulation and deviation, preparing the data set that includes data dependency, data quality, and data integration with various sources. In addition, the modelling of this data is one of the challenges related to data pre-processing, regarding data cleanliness, categorical data/sequence. In real-life applications, the following are common data problems:

- lack of metadata
- missing values
- data granularity
- integration data from multiple sources
- shortage of diverse samples
- design and management of the database, data lake
- quality of training data vs. real data

One study has highlighted the importance of data dependency, and states that data dependencies cost more than code dependencies in AI/ML systems, i.e. unstable or underutilized data dependencies (Sculley et al., 2015). Another issue mentioned is data drift, meaning that the statistical properties of predicting variables changing in an unforeseen way (Lwakatare et al., 2019; Munappy et al., 2019). Handling of data drifts in uploaded data, invalidation of models, e.g., due to changes in data sources, and the need to monitor models in production for staleness are problems mentioned in Lwakatare et al. (2019).

**Take away 5: AI development processes need to integrate infrastructures, processes and tools for managing data as their integral parts. It is not AI software, but AI data and software engineering.**

*4.4.4. Model development*

The list of the challenges regarding the building of the ML model is related to different aspects, such as a large number of model inputs, the AI ethics implementation, formal representation of complex models, optimization of feature engineering, imperfection and accuracy assurance, invalidation of models, model localization with data constraints, module documentation, uncertainty in input-output relationships, and uncertainty in model behaviour.

In model development, the main challenge is to obtain a large number of model inputs (Ishikawa, 2019; Renggli et al., 2019). It is difficult to clearly define the correction criteria for system outputs or correct outputs for each individual model input. Furthermore, for systems with a supervised learning paradigm, it is difficult to obtain labelled data that will serve as an input for the model, mainly when there is a large volume of unlabelled data. For a supervised learning paradigm, all samples must be labelled before the continuous integration service starts to work.

#### *4.4.5. Project management*

Project Management area competence (Alshangiti et al., 2019; Ishikawa, 2019; Martin and Schreckenghost, 2004; Selsam et al., 2017) and communication (Ishikawa, 2019; Vakkuri and Kemell, 2019) are challenges that stand out. As for competence, the main challenges are in resource management to meet the different platforms and technologies, knowledge of statistics and mathematics (need for specialization in these areas) (Ishikawa, 2019; Selsam et al., 2017), and skills required in different ML domains. For instance, the problem formulation phase requires a good overall understanding of how ML techniques work in order to help transform the problem in hand into an appropriate ML learning task. Model development requires hands-on and practical knowledge of tools, libraries and algorithms to build models (Alshangiti et al., 2019). Model fitting and tuning requires a solid understanding of the mathematical and statistical intuition behind the models (Alshangiti et al., 2019).

Communicating technical decisions and rationale during AI/ML model development is a non-trial task. Customers, even one with a technical background, do not often understand the technical details of AI models. It is challenging when one needs to explain the model functions to business stakeholders, which makes it difficult for the client to make decisions and evaluate the results (Ishikawa, 2019). Besides, communication of process, practice, and ethical regulation within the technical team is also overlooked. The direct consequence of this is the additional difficulty in estimating and assigning time and resources for the AI/ML parts (Munappy et al., 2019).

**Take away 6: AI project managers need competence and knowledge to act as a boundary-spanning role across AI/ML and non-AI/ML worlds**

#### *4.4.6. Infrastructure*

Infrastructure is a practical consideration when building and operating ML models. The challenges are related to the acquisition, installation, configuration, and maintenance of the necessary infrastructure for the development and operation of AI/ML Systems. One of the main engineering challenges perceived from managers was difficulty in building infrastructure (Lwakatare et al., 2019), and include the following:

- Hardware: whether local infrastructure or a cloud resource is used for each task of the ML development pipeline
- Platform: operating systems and installed packages. It might be a requirement that multiple platforms need to be executed simultaneously
- Tools: end-to-end pipeline supports include adoption of existing AI tools.

Tools are the most mentioned item in the category Infrastructure. Kim et al. (2018) state that "*despite a large suite of diverse tools, it is difficult for data scientists to access the right tools, because generic tools do not work for the specific problems that they have. It difficult to stay current with evolving tools, as they have other responsibilities and occasionally engage in data science work.*" Literature also reveals the lack of logging and tracking tools for AI development processes. As seen in Hilllaz et al. (2016), AI developers tend to track workflow via informal methods, such as emails among colleagues or notes to themselves. Complex and poor logging mechanisms during AI experiments are also stated, as seen in Lwakatare et al. (2019) and Hilllaz et al. (2016).

*4.4.7. Requirement Engineering*

Requirement Engineering (RE) challenges refer to understanding customer needs, business scenarios, obtaining requirements, specification, and data requirements (Nguyen-Duc, A., Abrahamsson, 2020).

Xie et al. (2011) state that the development of ML solutions has inherent complexities since it requires not only understanding what ML can and cannot do for organizations, but also specifying a well-defined business case and problem, translating and decomposing it into ML problem(s), data preparation and feature selection, ML algorithm selection and trade-offs, as well as finding linkages between ML models and business processes, among others. As mentioned in Rahman's study (Rahman et al., 2019), the requirements for the ML models are expected to be dynamic in nature and adapt to the rapidly evolving user requirements and business cases. Currently, the process and tool support for requirement engineering, particularly specification and validation of requirements for AI/ML software, is very limited (Kim et al., 2018). These RE challenges are also observed in SMEs and start-up contexts (Nguyen-Duc et al., 2020).

Similarly to traditional software projects, there can also be conflicting requirements where different teams with different experiences lead to conflicting goals (Mattos et al., 2017). In a supply chain system, there is an undeclared customer, as mentioned in Sculley et al. (2015), who silently uses the output of a given model as an input to another system. In classical software engineering, this is referred to as visibility debt.

**Take away 7: AI/ML system development needs new engineering guidance to identify, describe, analyze and manage AI software quality requirements.**

*4.4.8. Architecture Design*

Challenges are related to the architectural aspects of ML models, AI software or that which covers the entire developed system. Sculley et al. (2015) list the following risk factors when designing AI/ML systems: boundary erosion, entanglement, hidden feedback loops, undeclared consumers, data dependencies, configuration issues, changes in the external world. These authors (Sculley et al., 2015) also emphasize the concern of entanglement with the CACE principle of "changing anything changes everything". From a system perspective, the design of AI/ML software needs to be flexible in order to accommodate the rapidly evolving ML components (Rahman et al., 2019). Lwakatare et al. (2019) describe a specific type of design trade-off in customization of AI/ML platform functionalities.

*4.4.9. Model deployment*

Challenges regarding the deployment of the ML model in real or test environments involve dependency management, maintaining the glue code, monitoring and logging, and the unintended feedback loops. For the deployment process, when deploying the trained models from a testing environment to an operating one, there lacks a benchmarking understanding of the migration and quantization processes, such as the impacts on prediction accuracy and performance (Guo et al., 2019). Relating to the deployment process, changing hardware and software, issues to maintain reproducible results, incur engineering costs for keeping software and hardware up to date (Munappy et al., 2019).

*4.4.10. Integration*

The integration challenge occurs when integrating AI/ML modules to a larger system (Amershi et al., 2019; Tamir and Kandel, 1995), which impacts other engineering activities, such as deployment, testing, and maintenance.

#### *4.4.11. AI Engineering*

Challenges involve engineering models, processes, and practices when developing AI products, such as automation, lack of unifying different processes, lack of defined development process, lack of patterns, and the need for continuous engineering.

**Take away 8: The major challenges professionals face in the development of AI/ML systems include: testing, AI software quality, data management, model development, project management, infrastructure, and requirement engineering.**

### *4.5. RQ4. How are SE practices adjusted to deal with ML specific engineering challenges?*

To answer RQ4, we also performed a qualitative analysis of the citations referenced in the papers on SE practices applied in AI/ML development. Based on the information included in the extraction form, all practices and solutions cited in the papers were included in a single list. Then, each item of the practice was read and codified. We also classified items by area of knowledge (RQ5) to contextualize in which area the practice can be applied during the development of AI/ML systems. Finally, the context in which the practice was applied within the organizational context was included. It is important to note that two researchers with extensive experience in experimental software engineering and ML systems development evaluated and reviewed all codified practices. Nevertheless, we evaluated and revised the context in which the practice of SE is applied and the classification of the related areas of knowledge.

Table C.19 lists the 51 practices identified, the name of the codified practice, the references, the context, and the areas of knowledge that apply to SE. Regarding the context in which the practice was applied, this was classified into three groups: SI - Software Industry, Acad. - Academic, NA - Not Applicable, meaning the practice was applied in an experimental context or simulation.

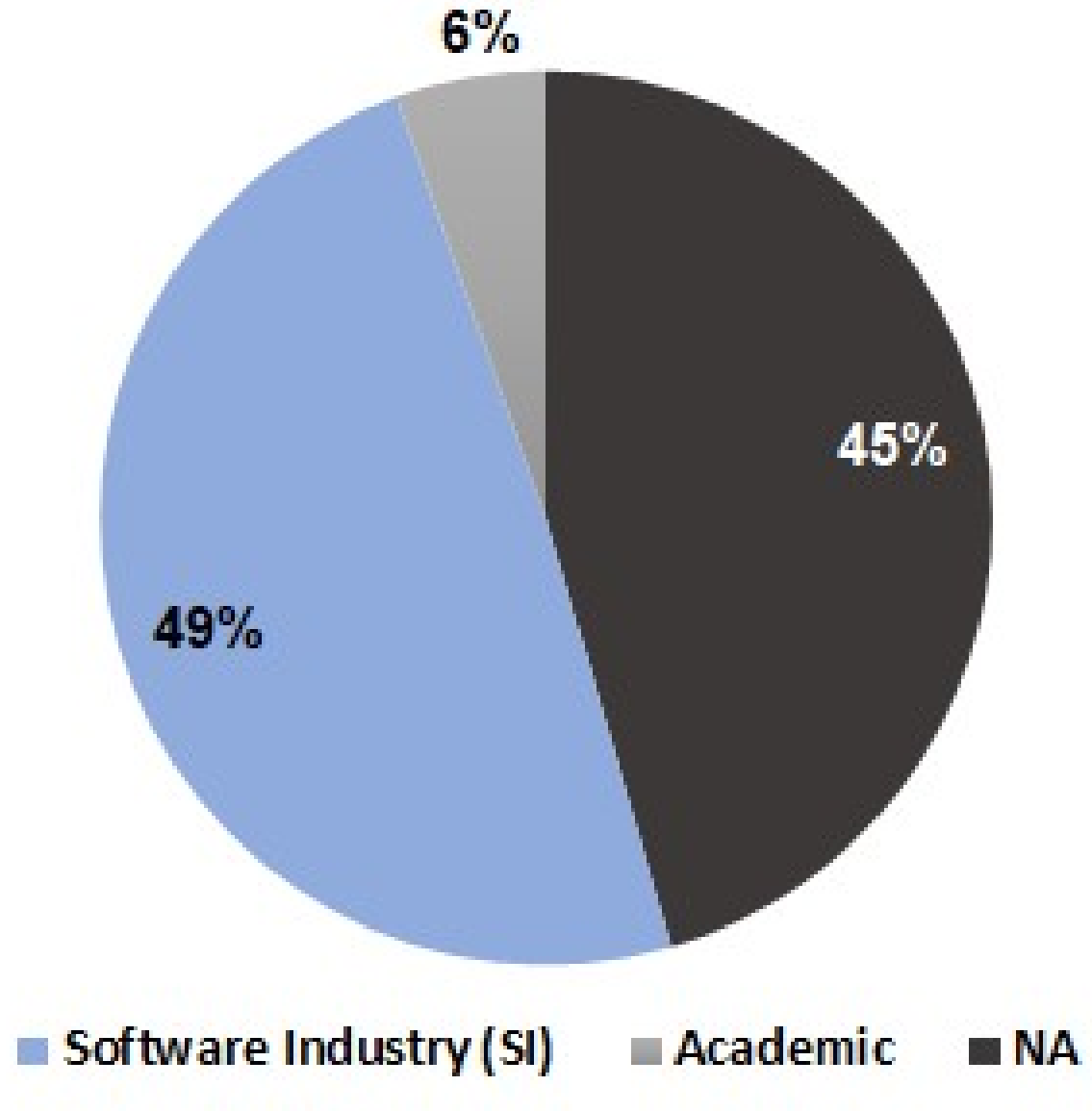

**Fig. 11.** SE practices by their application context.

Fig. 11 shows the context in which the SE practices are applied. In the graph on the left, 45% - NA (25 of the cases) are practices that have been applied in experimental environments, or are controlled studies that show initial results of proposals that have not yet been applied in the industry. However, 49% (27 of the cases) are practices applied in the software industry and 6% (3 of the cases) are applied in an academic environment.

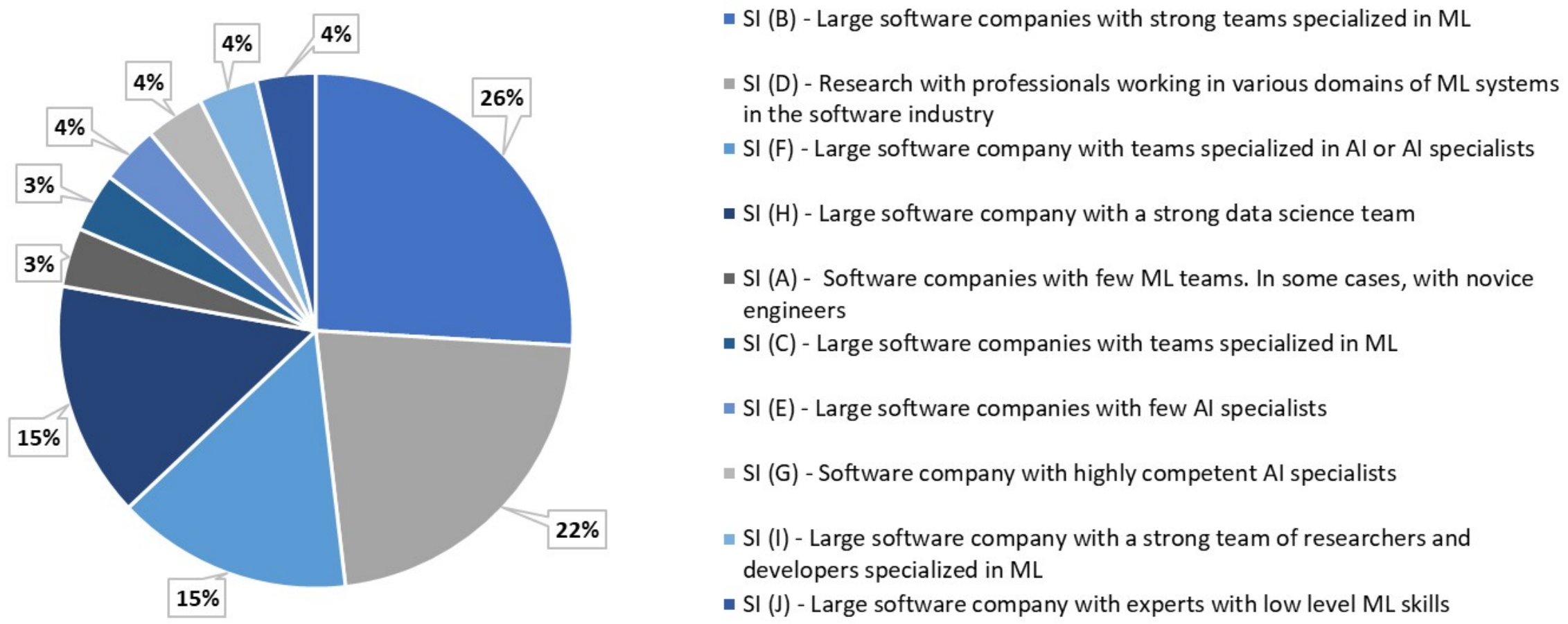

**Fig. 12.** SE practices by their software industry context.

In greater detail, Fig. 12 shows in which context of the 45% of the software industry the practices were developed. As a result, we can describe the level of AI/ML knowledge of the team or specialists who developed these practices:

SI (A). Software companies with few ML teams. In some cases, with novice engineers.
SI (B). Large software companies with strong teams specialized in ML.
SI (C). Large software companies with teams specialized in ML.
SI (D). Research with professionals working in various domains of ML systems in the software industry.
SI (E). Large software companies with few AI specialists.
SI (F). Large software company with teams specialized in AI or AI specialists.
SI (G). Software company with highly competent AI specialists.
SI (H). Large software company with a strong data science team.
SI (I). Large software company with a strong team of researchers and developers specialized in ML.
SI (J). Large software company with experts with low level ML skills

Moreover, regarding the practices applied in the industrial environment, in the graph on the right, 26% SI (B) are applied in large companies by teams specialized in ML. And 22% SI (D) are practices obtained through research with professionals who work in various domains of ML systems in the software industry.

In the context of the software industry, different practices are applied, such as those of continuous integration (Renggli et al., 2019) and experimentation (Lwakatare et al., 2019), which are important practices for the data pipeline of AI/ML projects and support data management, reuse, automation, and traceability (Mattos et al., 2017; Amershi et al., 2019). For example, Amershi et al. (2019) mention that the teams built an architecture integrated with the projects to automate the workflow from end to end in the data pipeline. In Lwakatare et al. (2019), the team developed an online experimentation platform for the development of ML models for their projects. Renggli et al. (2019) created a tool called Ease.ML/CI in order to apply continuous integration in projects with ML. Mattos et al. (2017) developed a framework to support the architecture of the project by providing continuous and automated experimentation.

Furthermore, practices that support the implementation of ethics in AI/ML projects (Bryson and Winfield, 2017; Vakkuri et al., 2019; Vakkuri and Kemell, 2019) are important to address aspects such as transparency, accountability, responsibility, and fairness in the development of AI-based software. Vakkuri et al. (2019) investigated the practice of AI ethics in healthcare sector solutions and assessed how AI ethics is implemented. The authors found that current practices in the field are already capable of producing transparency in AI systems development.

In addition, we found some practices that can support data validation during the construction of the solution, and which are important for improving data quality for ML projects (Kim et al., 2018; Rahman et al., 2019). For example, Kim et al. (2018) verified which practices data scientists use to improve the quality of project data, and the main data validation methods applied by professionals are the following: cross-validation is multi-dimensional, check data distribution, dogfood simulation, type, and schema checking, repeatability, and check implicit constraints. Another example can be found in Rahman et al. (2019), who shared the team's experience in developing ML-based automatic detection and error correction in retail transactions. The recommendations for machine learning are applied at different stages of development, for example, in the pre-processing stage of the data, in which the team fills in the missing values and normalizes the data fields to correct formatting differences, and thus improve the data quality.

Additionally, testing practices such as stress (Chakravarty, 2010) and combinatorial (Munappy et al., 2019) tests are applied in the industry for systems based on AI/ML. For example, Munappy et al. (2019) applied a combinatorial test for AI based on data, and Chakravarty (2010) applied a stress test to measure end-to-end performance in AI-based web services.

**Take away 9: The laboratory (experimental or simulation) context is where practices are mostly applied. Second is the software industry context, where practices are mostly applied in large companies by teams specialized in ML.**

*4.6. RQ5. What are the SE practices reported in the primary studies? (SWEBOK)*

Regarding the SE knowledge areas that the practices apply to, Fig. 13 **Error! Reference source not found.** shows three areas of knowledge with greater coverage. The area for Testing presents 46% of the cases, Design, and Configuration Management presents 42% in each area. The areas of Construction (26%), Professional Practice (21%), Requirements and Process (18%) are areas that also show technologies and/or practices that are being developed to meet these areas. The areas with the lowest number of SE for AI/ML practices were Deployment, Maintenance, Quality, and Management.

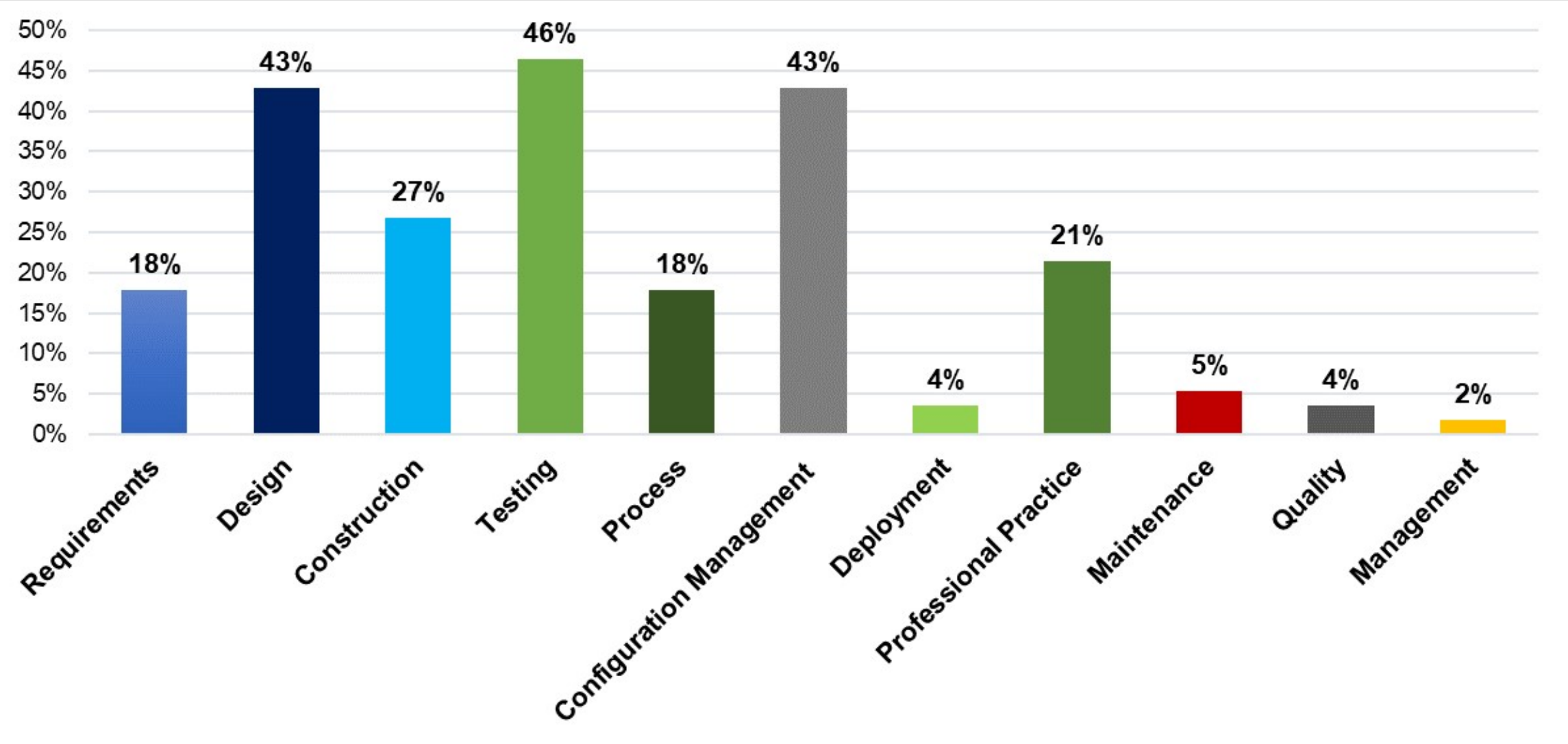

**Fig. 13.** SWEBOK knowledge area associated with AI software.

When we compare the adoption of practices with the engineering challenges mentioned in the list of challenges in Table 16 (Subsection 5.2), we note that there are challenges supported by SE practices with different contribution types as mentioned by Berg et al. (2018) tools, platforms, approaches, guidelines to support the AI/ML system development process. For example, in the area of Architecture, Component Reuse was applied to the development of projects based on Artificial Neural Networks (ANN) (Ghofrani et al., 2019).

For example, for Tests, bug detection practices in ML systems (Guo et al., 2019; Thung et al., 2012) have been adjusted to identify errors in ML systems and on Deep Learning (DL) platforms. In the area of configuration management, Lwakatare et al. (2019), Mattos et al. (2017) and Renggli et al. (2019) proposed the creation of experimentation environments and automation/continuous integration. In addition, different guidelines proposed by Rahman et al. (2019), Amershi et al. (2019), Kim et al. (2018), Martin and Schreckenghost (2004), Hilllaz et al. (2016), Martin (2016), and Hannay et al. (2009) were tried by professionals in real ML/AI projects. These practices include guidelines for reuse, data automation, and data traceability to support configuration management.

In the area of **Software Quality** practices to meet AI ethics, the authors cited existing proposals to support ethics in AI projects in planning and design, such as ART principles (Dignum, 2019), Guidelines for Ethically Aligned Design (How, 2018), Ethics Guidelines for Trustworthy AI (High Level Independent Group on Artificial Intelligence (AI HLEG), 2019). In addition to these, we found new proposals, such as practices to meet some aspects of AI ethics (Vakkuri et al., 2019) that are based on practices developed by professionals working in the area. The practice for ETHICS IN AI - are ethical considerations in Artificial Intelligence and Autonomous Systems (Bryson and Winfield, 2017), which gathers compliance standards regarding the confidence in the effectiveness of a system in important areas for users, such as safety, protection, and reliability. It is an approach called 'RESOLVEDD' and is strategic to project the AI ethics (Vakkuri and Kemell, 2019) defining aspects, such as the commitment to the ethically aligned project, transparency in design, and responsibility in design.

For **Testing**, SE practices were directed to give specific Support in ML systems Tests, General Techniques/Guidelines that include practices for ML systems tests, bug detection in ML systems, stress test, and specific tests for neural networks/Deep Learning. Table 12 details the proposed practices for the software testing area.

**Table 12.** SE approaches, processes, practices, or tools reported for AI Testing.

| **Proposed SE practices for Testing** | **Practice description** | **Mentioned in** |
|---|---|---|
| Approaches for supporting ML system testing | Methodology for automatic data entry generation for tests in ML systems | Aggarwal et al. (2019) |
| | Combinatorial tests to gain insights into the quality of ML solutions | Barash et al. (2019) |
| | Tests on the implementation of machine learning classification algorithms | Xie et al. (2011) |
| General guidelines (guidelines that include testing practices) | Suggestions for support in difficulties in the engineering of ML-based systems | Ishikawa (2019) |
| | Guidelines used by professionals as practices for SE specialized in ML | Hilllaz et al. (2016) |
| | SE practices for scientists who work and develop their daily work using scientific software | Hannay et al. (2009) |
| | Best practices for improving and ensuring the quality of work for data scientists | Kim et al. (2018) |
| | Guidelines for researchers and best practice | Rahman et |
| | practitioners in ML application development | al. (2019) |
| | Best practice guidelines for ML | Martin (2016) |
| Bug detection | Guidelines for bug detection for evaluating new DL structures and platforms | Guo et al. (2019) |
| | Guidelines for investigating errors in ML systems | Thung et al. (2012) |
| | Ease.ML/CI is a tool for supporting the software engineering lifecycle, and for making continuous integration in ML systems, including the testing stage | Renggli et al. (2019) |
| | Interactive proofing assistant to support the implementation of ML systems, including bug identification | Selsam et al. (2017) |
| Stress testing | Stress testing to measure end-to-end performance on AI-based Web Services | Chakravarty (2010) |
| Specific tests for Neural Networks/Deep learning | DNN operational testing to determine the model's actual performance in specific operating contexts | Li et al. (2019) |
| | Reproducible test to ensure the robustness of a model (CleverHans) | Goodfellow and Papernot (2017) |
| | Metamorphic testing, a deep learning framework validation approach for automated images | DIng et al. (2017) |
| | White box testing for large-scale DL systems (DeepXplore) | Pei et al. (2019) |
| | Multiple granularity test criteria set for DL systems (DeepGauge) | L. Ma et al. (2018) |
| | A resource-driven approach to test the resilience of image classification networks against contradictory examples | Wicker et al. (2018) |
| | Systematic testing to analyze autonomous DNN-based steering systems | Zhang et al. (2018) |
| | A technique for debugging neural network models (MODE) | S. Ma et al. (2018) |
| | An economical testing approach for DNNs (DeepConcolic) | Sun et al. (2018b) |
| | White box test methodology for DNNs based on the MC/DC test criteria | Sun et al. (2018a) |
| | Systematically tests different parts of DNN logic by generating test inputs that maximize the number of activated neurons (DeepTest) | Jana et al. (2018) |

In **Configuration Management**, most ES practices are general guidelines that include practices for configuration management (Table 13). However, specific practices were also proposed, such as providing environments to support the development team (Lwakatare et al., 2019; Selsam et al., 2017) providing environments for experimentation and automation/continuous integration (Lwakatare et al., 2019; Mattos et al., 2017; Renggli et al., 2019), reuse (Ghofrani et al., 2019), stability (Yokoyama, 2019) and strategies developed for specific domains, such as multi-agent systems (Damasceno et al., 2018), systems for automated steering (Chen et al., 2018), and systems with DL components (Arpteg et al., 2018).

**Table 13.** SE approaches, processes, practices, or tools reported for AI configuration management.

| SE practices for configuration management | Practice description | Mentioned in |
|---|---|---|
| Experimentation, automation, and continuous integration environments | An architecture framework to provide automated continuous experimentation | Mattos et al. (2017) |
| | Online experimentation platform – An online experimentation platform that consists mainly of four components: experimentation portal, experimentation execution service, log processing service, and analysis service. It also allows the entire team to use the platform to experiment with the system under development. | Lwakatare et al. (2019) |
| | Ease.ML/CI - a continuous integration system for ML - the tool is designed to support the automation and continuous integration of ML components | Renggli et al. (2019) |
| Platforms for automated development | Practices for automated driving software development. These are practices intended to interpret sensor data in order to understand the information contained at a high level. | Lwakatare et al. (2019) |
| | A platform for automating information extraction in order to optimize communication between team and stakeholders. A platform designed to label and form the validation dataset for building ML models and automating information extraction from the databases. | Lwakatare et al. (2019) |
| | An AI platform for supporting data science teams. A web platform to communicate requirements to the product development team, provide internal AI training and carry out projects with external companies | Lwakatare et al. (2019) |
| | An interactive proof assistant for support in the implementation of the ML system - An interactive proof assistant to both implement their ML system and to state a formal theorem defining what it means for their ML system to be correct. | Selsam et al. (2017) |
| Strategies developed for specific domains | A framework to support AI Ethics - RESOLVEDD Strategy. Framework RESOLVEDD strategy considers ethical aspects, such as commitment to ethically aligned design, transparency in design, accountability in design, and responsibility in design. | Vakkuri and Kemell (2019) |
| | A methodology framework for safety validation of automated driving systems. A methodology based in Causal Scenario Analysis (CSA) has been introduced as a novel systematic paradigm to derive and evaluate the scenarios in automated driving systems. | Chen et al. (2018) |
| | Guidelines of Metric-Based Evaluation of Multiagent Systems Models. This proposal includes quality dimensions for multi-agent systems (MAS), such as model expectations, resources quality, reliability, methodology adequacy, maintainability, and agent quality. | Damasceno et al. (2018) |
| | A model for KnowLang: A Language for knowledge representation in autonomic service-component ensembles - The logical framework provides additional support for the domain ontology computational structures that determine the logical foundations helping a service component (SC) reason and infer knowledge. | Vassev et al. (2011) |
| Reuse | Lessons learned to be applied in reusability methods in the development of ANN-based systems. | Ghofrani et al. (2019) |
| Stability | An architectural pattern to improve the operational stability of ML systems. This architectural pattern helps the operators break down the failures into a business logic part and an ML-specific part. | Yokoyama (2019) |

In the **Design** area, most practices are specific for supporting the design of ML systems, such as the proposal of an architectural structure to provide automated continuous experimentation (Mattos et al., 2017). New architectural standards have also been proposed to improve the operational stability of ML systems (Yokoyama, 2019). Additionally, there are architectural proposals to design specific AI systems, such as the proposal for an architectural structure to facilitate the design and implementation of hybrid intelligent forecasting applications (Lertpalangsunti and Chan, 1998). Table 14 details the proposed practices for the software design area.

**Table 14.** SE approaches, processes, practices, or tools reported for AI design.

| Proposed SE practices for design | Practice description | Mentioned in |
|---|---|---|
| Framework for continuous experimentation | An architecture framework to provide automated continuous experimentation. The framework supports automated experimentation to ensure architectural qualities, such as external adaptation control, data collection, performance, explicit representation of the learning component, decentralized adaptation, and knowledge exchange. | Mattos et al. (2017) |
| Architectural patterns | A model for architecture and design (anti-)patterns for Machine Learning systems. A model of architectural patterns was proposed for ML systems with patterns that apply to many stages of the pipeline or many phases of the development process. | Washizaki et al. (2019) |
| | A framework of anti-patterns for technical debt in ML systems. A set of anti-patterns and practices for avoiding technical debt in systems using ML components are presented. | Sculley et al. (2015) |
| | An architectural pattern to improve the operational stability of ML systems. This architectural pattern helps the operators break down the failures into a business logic part and an ML-specific part. | Yokoyama (2019) |
| | A solution for representing generic and well-proven ML designs for commonly-known and recurring business analytics problems | Nalchigar et al. (2019) |
| Strategies for Designing Ethics | Guidelines for implementing ethics in AI presented for aspects from the ethical to the development process, such as ethics in design, and ethics for design | Vakkuri et al. (2019) |
| | The Framework RESOLVEDD strategy considers ethical aspects, such as commitment to ethically aligned design, transparency in design, accountability in design, and responsibility in design. | Vakkuri and Kemell (2019) |
| | ETHICS IN AI - Ethical Considerations for Artificial Intelligence and autonomous systems | Bryson and Winfield (2017) |
| Strategies for designing Security | A methodology framework toward safety validation for automated driving systems. A methodology based on causal scenario analysis (CSA) has been introduced as a novel systematic paradigm for deriving and evaluating the scenarios in automated driving systems. | Chen et al. (2018) |
| | Four different strategies for achieving safety in the machine learning context. Strategies for achieving safety in ML projects for defining safety in terms of risk, epistemic uncertainty, and the harm incurred by unwanted outcomes. | Varshney (2017) |
| Design specific systems | Guidelines of Metric-Based Evaluation of Multiagent Systems Models. This proposal includes quality dimensions for multiagent systems (MAS), such as model expectations, resources quality, reliability, methodology adequacy, maintainability, and agent quality. | Damasceno et al. (2018) |
| | A model for KnowLang: A language for knowledge representation in autonomic service-component ensembles. The logical framework provides additional support for the computational structures of domain ontology that determine the logical foundations that help a service component (SC) reason and infer knowledge. | Vassev et al. (2011) |
| | A Framework for agent-based hybrid intelligent systems (HIS). An architectural framework that overcomes the technical impediments and facilitates hybrid intelligent system construction is presented. | Li and Li (2011) |
| | An architectural framework to facilitate the design and implementation of hybrid intelligent forecasting applications. | Lertpalangsunti and Chan (1998) |
| General guidelines that include design practices | Checklists for development processes in ML systems support: business modeling and data processing. Checklist to support data Processing (CheckDP) - this was developed to support the structuring of the data task. In this stage, the developer performs different subtasks (e.g., verifying that the data is complete). | Nascimento et al. (2019) |
| | Lessons learned from the AI/ML projects at Microsoft. The best practices set are viewpoints on some of the essential challenges associated with building large-scale ML applications and platforms, and how the teams address them in their products. | Amershi et al. (2019) |
| | Map of SE challenges for building systems with DL | Arpteg et al. |
| | components. A set of 12 challenges were identified and these outline the main SE challenges involved in building systems with DL components that were described for the areas of development, production, and organizational challenges. | (2018) |
| | Best Practices to improve and ensure the quality of data scientists work. The best practices to overcome the challenge of data science work and advice that a data scientist "expert" would give to novice data scientists. | Kim et al. (2018) |
| | Guidelines for researchers and practitioners for best practices in the development of ML applications. Guidelines for researchers and practitioners with best practices for the development of ML applications, with three distinct perspectives: software engineering, machine learning, and industry-academia collaboration. | Rahman et al. (2019) |
| | Guidelines of the best practices for machine learning. A set of 43 best practices for machine learning from Google. These best practices serve as guides together with other popular guides to practical programming. | Martin (2016) |

**Take away 10: The three areas of SWEBOK with the most proposed practices for supporting the development of AI/ML systems are testing, design, and configuration management.**

Additionally, we classified the contribution type of each practice (primary study) based on Paternoster et al. (2014). According to Paternoster et al. (2014), the contribution types these include are for (1) model, (2) theory, (3) framework/methods, (4) guidelines, (5) lessons learned, (6) advice/implications, and (7) tools, as shown in Table 15.

**Table 15.** Contribution type to the practice based on Paternoster et al. (2014).

| Contribution Type | Description |
|---|---|
| **Model** | Representation of an observed reality by concepts or related concepts after a conceptualization process |
| **Theory** | Construction of cause-effect relationships from determining results |
| **Framework/methods** | Models related to constructing software or managing development processes |
| **Guidelines** | List of advice, synthesis of the obtained research results |
| **Lessons learned** | Set of outcomes directly analyzed from the obtained research result |
| **Advice/implications** | Discursive and generic recommendations from personal opinions |
| **Tool** | Type of technology, program, or application used to create, debug, maintain or support development processes |

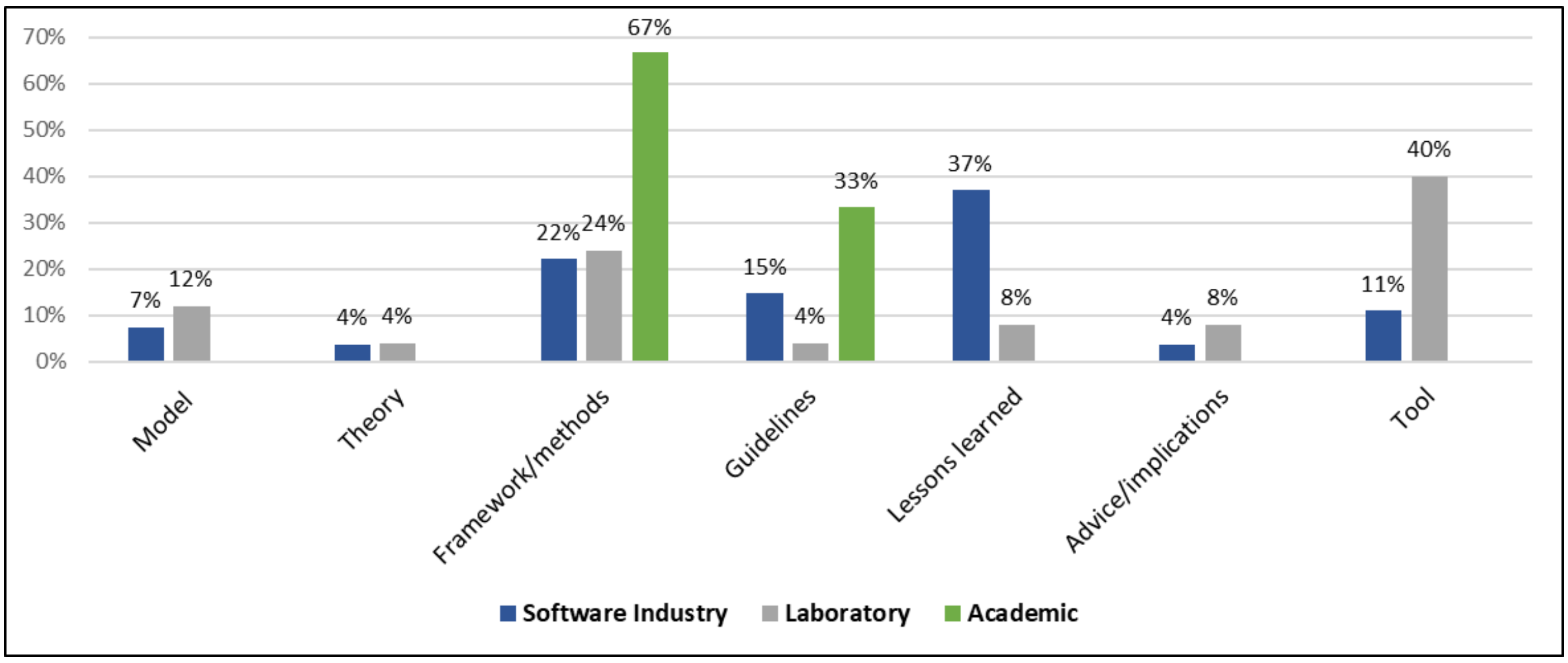

**Fig. 14.** Contribution types proposed for each context.

Fig. 14 illustrates contribution types as proposed by each context: software industry, academic, and laboratory, in other words, when or whether the practice was applied in an experimental or simulation context. In the context of the software industry, the lessons learned are the most frequently proposed contribution type (37%), followed by the framework/methods (22%), and guidelines (15%). In the laboratory context, the most frequently proposed contribution type is tools (40%), followed by the framework/methods (24%), and model (12%). In the academic context, framework/methods (67%) is more frequently proposed.

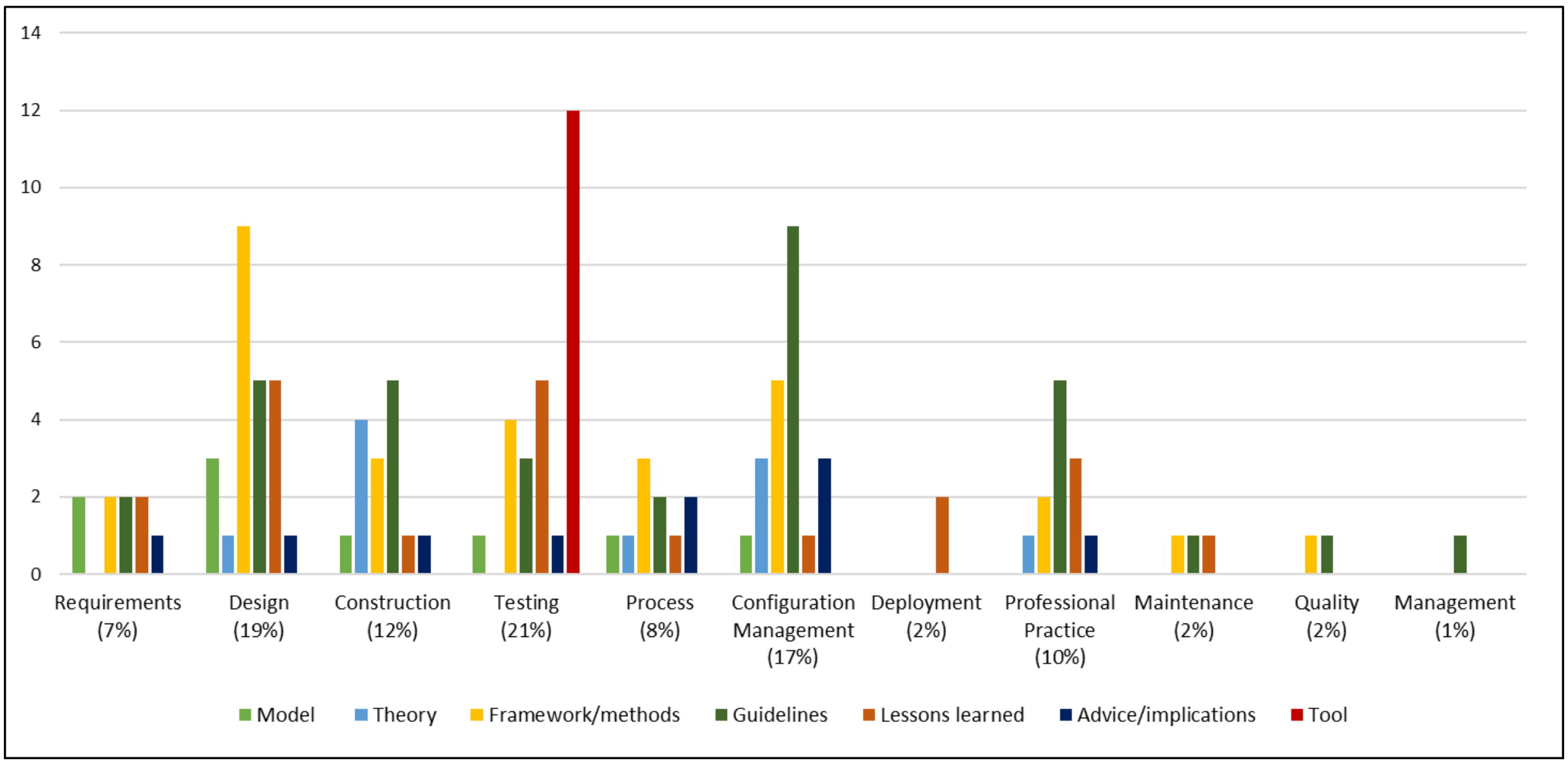

**Fig. 15.** Contribution types proposed for each SWEBOK knowledge area.

Fig. 15 illustrates the proposed contribution types for each SWEBOK knowledge area. The 11 knowledge areas represent a total of 126 occurrances through 7 different contribution types. These include (1) model, (2) theory, (3) framework, (4) guidelines, (5) lessons learned, (6) advice, and (7) tools. The SWEBOK knowledge areas that presented the most contribution types were testing (21%), design (19%), and configuration management (17%), followed by construction (12%) and professional practice (10%).

For example, in the testing area, the tools are the most often proposed contribution type (46%), followed by the lessons learned (19%), and framework/methods (15%). In the design area, the framework/methods are the most proposed contribution type (38%), followed by the lessons learned (21%), and guidelines (21%). While in the configuration management area, the guidelines (41%), followed by the framework/methods (21%), theory, and advice (14%). In the construction areas, we most often find guidelines (33%), theory (27%), and framework (20%). In professional practice, guidelines (42%), lessons learned (25%), and framework (21%) are the most cited.

**Take away 11: The contribution types of the most frequently proposed SE practices for supporting the development of AI/ML systems are framework/methods, guidelines, lessons learned, and tools.**

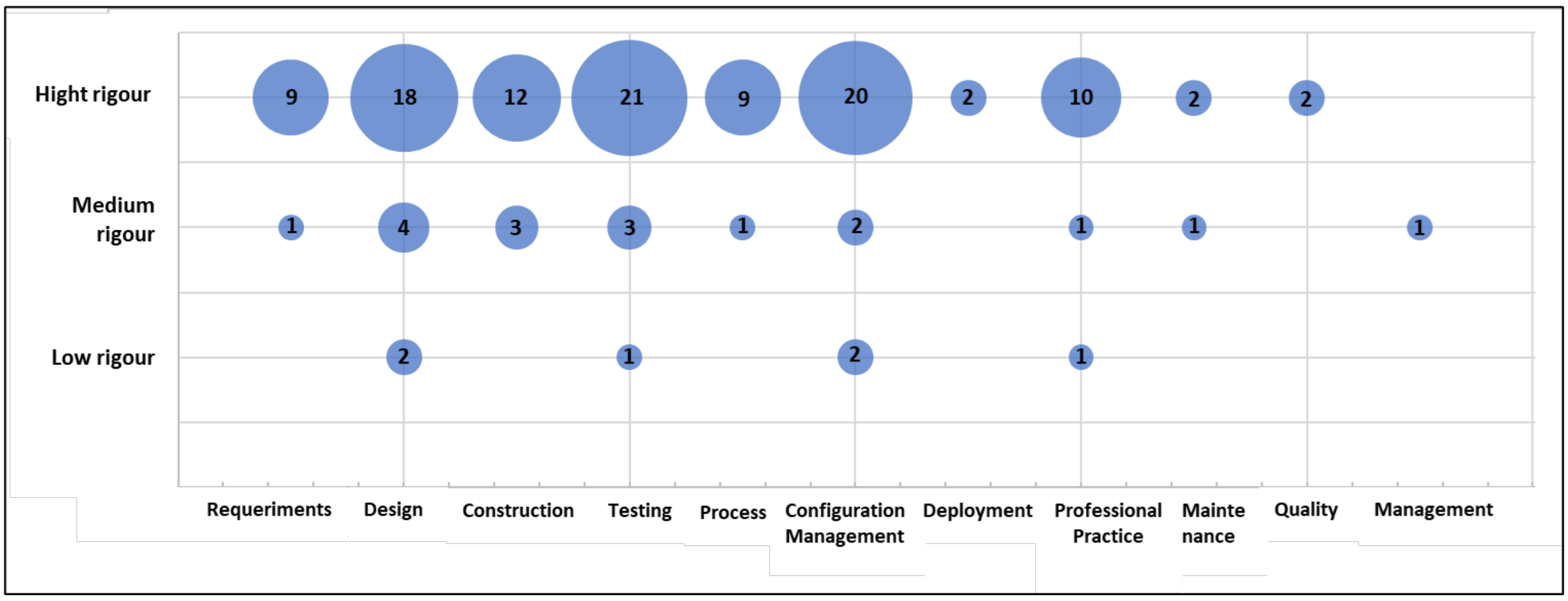

**Fig. 16.** Rigour of each covered knowledge area, 1990–2019.

Fig. 16 shows the degree of rigour within each knowledge area between 1990–2019. The x-axis represents the knowledge areas, while the y- axis represents the rigour. Four knowledge areas had papers that received low rigour score (Braiek and Khomh, 2020; DIng et al., 2017; Varshney, 2017; Yokoyama, 2019), as they did not provide enough details about the research design, sampling, data collection, data analysis and included no assessment of the validity of the results.

### *4.7. RQ6. Which type of empirical methods are used?*

Regarding the experimental methods used in primary studies, we classified them into two categories: (a) the research method and (b) the research strategy, as shown in Fig. 17. According to Storey et al. (2020), a research strategy is a broader term that may involve the use of one or more methods for collecting data, and a research method is a technique used to collect data. Wohlin et al. (2012) describe the research methods that are frequently applied in software engineering such as surveys, case studies, and experiments. For the research strategies we used the classification by Stol and Fitzgerald (2018) that are field study, field experiment, experiment simulation, laboratory experiment, judgment study, sample studies, formal theory, and computer simulation. For the research method, the results showed that 42% of the primary studies are case study, and 35% experiments, and 21% survey. As for the research strategy, 25% of the cases are field studies, 28% laboratory experiments, 9% field experiments, and 9% judgement study.

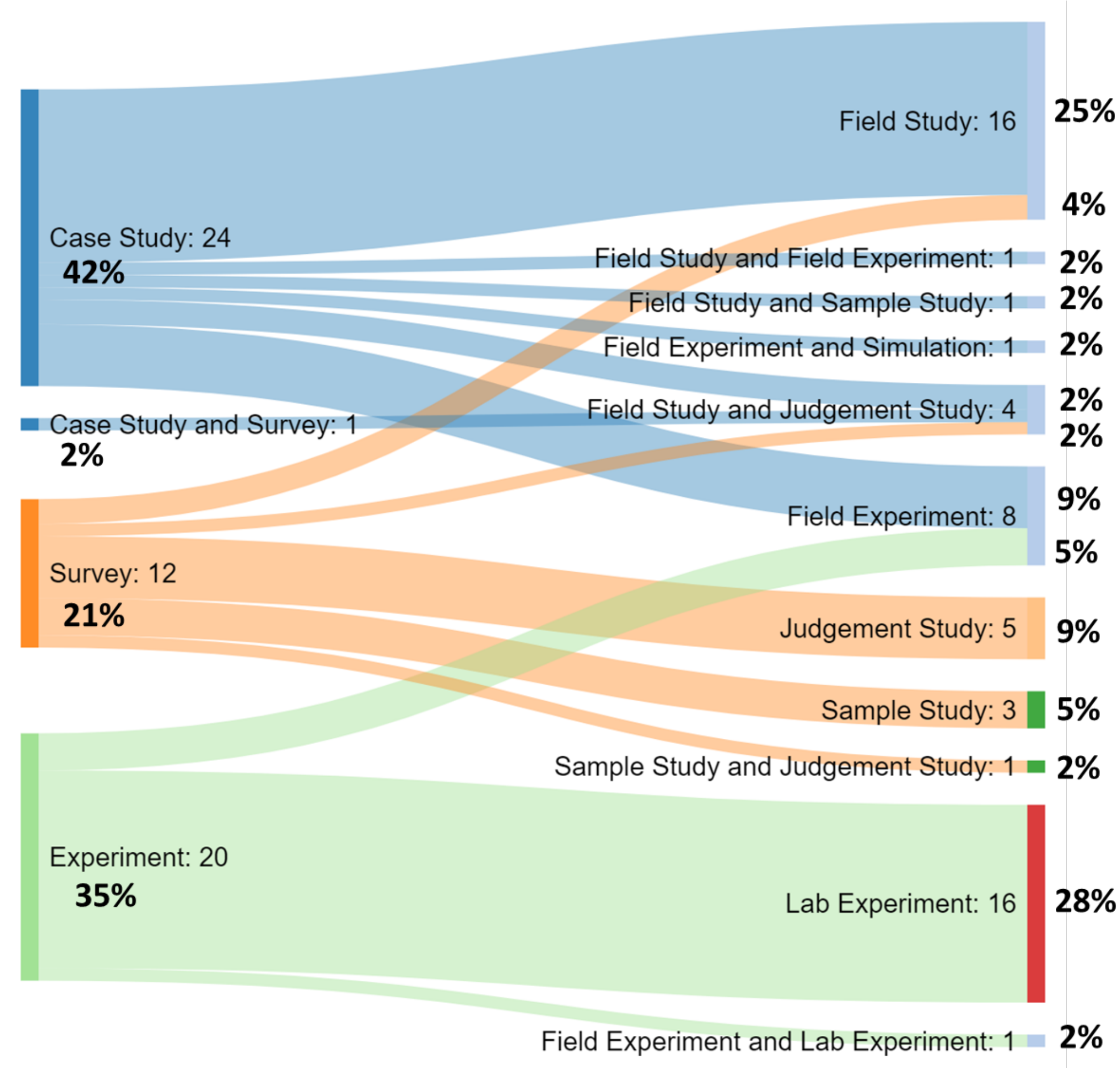

**Fig. 17.** Types of empirical methods used in primary studies.

**Take away 12: The types of empirical methods mostly used in research on AI/ML development are case studies, and the research strategies adopted are field studies and field experiments.**

## 5. Discussion

As a result of the SLR, we were able to identify the challenges and practices of SE, and, in turn, analyzed and summarized them in categories and subcategories. These two topics deserve further investigation and will be discussed in the following subsections.

### *5.1. Challenges faced by professionals working in the development of AI/ML systems*

Although classifying problems and practices using our current categories of project context quantitatively result in meaningful patterns of context-problem-practices, we did observe commonalities among several primary studies. Firstly, in our comparison of AI development challenges in different environmental contexts, i.e., among the laboratory environment (experimental), small and large companies present some interesting observations that are worth discussing further. The challenges in laboratories seem to mainly be associated with model building and testing phases only, for instance, lack of understanding of the structures for training, prediction, and robustness of trained models (Guo et al., 2019), and the issues of individual discrimination in ML models (Aggarwal et al., 2019), combinatorial tests to cover requirements (Barash et al., 2019), and characteristics of bugs in ML systems (Thung et al., 2012). Also, the test needs to produce accurate results to label the data collected in the field (Li et al., 2019). These problems may be related to a lack of understanding of the behaviour of ML models.

In start-ups and small companies, we found more frequently mentioned challenges regarding processes and practices, for instance, the lack of definition of the process to be followed by professionals (Nascimento et al., 2019). These results confirm the challenges highlighted in previous studies by Nguyen-Duc et al. (2020) and Nascimento et al. (2019). We realize that the problems faced by professionals in small companies are due to the lack of software processes to guide professionals in their activities, the lack of infrastructure for the project's data pipeline, mainly to support the integration of the components ML in legacy systems. Notwithstanding, we found no additional challenges to those already reported for this context.

For the context of large companies, the challenges are diverse and different from the context of small companies in some aspects, for example, AI projects in large companies include challenges related to AI ethics, tests in ML applications, scalable ML models, and team management. The challenges in the test are related to the generation of test cases that require balancing the amplitude of the test cases with a depth of response quality (Chakravarty, 2010). Moreover, there are challenges in managing large teams, with different cultures and in different countries, and with different knowledge. Our results confirm that the biggest challenges reported in the primary studies for the development of these systems are associated with data management, model development, software quality, testing, and infrastructure.

We also found some related works that present challenges similar to ours (Table 16). Bosch et al. (2020) present the challenge for data quality management similar to the challenges listed in the data management and AI software quality categories. In addition, the challenges for design methods and processes are similar to our challenges listed in requirement engineering, and project management. However, we have included challenges for the Testing and Architecture Design categories, for example. Zhang et al. (2020) present challenges related to ML testing similar to the challenges in our Testing category. Washizaki et al. (2019) present challenges related to architecture standards similar to the challenges of the Architecture Design category. Finally, Hutchins et al. (2017) present challenges related to AI ethics similar to the challenges of AI software quality.

### 5.2. *Practices support to challenges*

We mapped the challenges vs. practices proposed to support professionals in these challenges, and it is possible to perceive that there are open challenges that have not yet been filled, and that there are practices lacking, mainly in the areas of maintenance, quality, implementation, and project management in ML systems. Only 46% of the total number of practices were extracted in relation to an industrial context, for example, in large companies with teams specialized in ML in the areas of configuration management, project, professional practice. There is a significant number of practices (49%) proposed without empirical investigation in industrial environments. Interestingly, most of the testing practices and processes fall into this category. Most of these practices were proposed for specific tests on artificial neural networks/deep learning and evaluated in a controlled environment; therefore, the SLR results show that experimental evidence in a real environment (software industry) was lacking for most of the proposed practices in the testing area, and showed its usefulness and applicability. Our SLR is one of the first studies that presents this type of analysis on the challenges vs. practices, and below we present an overview of the practices proposed according to context and which challenges are associated.

We mapped the reported challenges and their context and analyzed the practices developed to overcome these challenges. In the context of academia, for instance, the main challenges are related to AI software quality, project management, infrastructure, and education. However, the proposed practices do not meet all challenges and are specific to software quality, as a framework to support the definition of AI ethical strategies (Lwakatare et al., 2019), and as guidelines to support the evaluation of multi-agent systems (Damasceno et al., 2018). Additionally, for the challenge in AI/ML education, only one proposal was found, which was a framework to support the teaching of data science (Zhang et al., 2017).

The challenges reported in the experimental laboratory environment are related to testing, AI software quality, data management, model development, project management, requirement engineering, ai engineering, and model deployment. However, most of the proposals relate to challenges in the testing area, and most are tools or frameworks for testing neural network systems or deep learning. To address AI software quality challenges, some solutions have been proposed, such as a deep learning validation framework using metamorphic tests (DIng et al., 2017), guidelines to support the definition of security strategies (Varshney, 2017), models to improve the operational stability of the ML systems (Yokoyama, 2019) and advice for the development of expert systems (Motoda, 1990). To overcome the challenges in the development of models, we found a proposal that presents a set of lessons learned for evaluating new DL structures and platforms and which can be useful in identifying bugs in ML systems (Guo et al., 2019). However, we have not found proposals for overcoming the challenges in project management, AI engineering, infrastructure, architecture design, integration, operational support, and education.

The challenges reported in small businesses are related to data management, requirements, architecture, and integration. The practices we found in primary studies for overcoming the requirements and architecture challenges were a checklist-based framework for supporting professionals during business modelling and data processing activities (Nascimento et al., 2019). The RUDE model was proposed to support prototyping also in the requirements area (Kouba et al., 1992), and the proposals we found specifically address some challenges subcategories of the requirements and architecture. There is also a proposal for meeting data management, integration, and other aspects of requirements, and architecture.

The challenges reported in large companies are diverse, and the main ones are AI software quality, data management, model development, infrastructure, ai engineering, architecture, and testing. Lessons learned are the most frequent proposals to overcome these challenges, followed by tools and guidelines (Table 16). For example, the lessons learned are based on the experiences of other professionals who reported the problems they experienced and how they solved these problems (Kim et al., 2018; Lwakatare et al., 2019; Amershi et al., 2019). The proposed tools are for supporting model experimentation (Lwakatare et al., 2019), and for continuous integration of ML systems (Renggli et al., 2019). However, there are few proposals to overcome the challenges of implementing the model, integration, and operational support.

**Table 16**. Challenges vs. practices in the context of the software industry.

| Challenges: | Type contributions of the Practices: | Mentioned in: | Related works: (Challenges) |
|---|---|---|---|
| Testing | Theory, Framework, Guidelines, Lessons learned, Tools | P3, P22, P29, P32, P45 | Zhang et al. (2020) |
| AI software quality | Theory, Guidelines, Lessons learned e Tools | P2, P3, P4, P5, P30, P42, P32, P33, | Bosch et al. (2020), Hutchins et al. (2017) |
| Data management | Framework, Lessons learned, Tools | P1, P2, P3, P5, P30, P34, P42 | Bosch et al. (2020) |
| Model | Framework, Guidelines, Lessons | P2, P3, P4, P5, | Bosch et al. |
| development | learned, Tools | P16, P28, P32, P42 | (2020) |
| Project management | Theory, Framework, Lessons learned, Tools | P2, P3, P5, P8, P13, P32, P33, P34, | - |
| Infrastructure | Theory, Framework, Lessons learned, Tools | P2, P3, P5, P10, P13, P33, P34, P42 | Bosch et al. (2020) |
| Requirement engineering | Framework, Lessons learned, | P1, P2, P3, P8, P16, P34, P42, P51, | Bosch et al. (2020) |
| AI engineering | Model, Framework, Lessons learned, Advice and Tools | P3, P8, P10, P29 | - |
| Architecture design | Framework, Guidelines, Lessons learned, | P1, P3, P5, P8, P34, P46, | Bosch et al. (2020), Washizaki et al. (2019) |
| Model deployment | Theory, Framework, Guidelines | P4, P33, P28, | Bosch et al. (2020) |
| Integration | Model, Lessons learned | P2, P14, | - |
| Operation support | Lessons learned | P2 | - |

### 5.3. *Threats to validity*

Although this study systematically followed the protocol according to Kitchenham and Keele (2007) to reduce the threats to the validity of the study, some decisions may have affected the results. In the following, threats to validity will be discussed based on the types presented in Wohlin et al. (2012) and Petersen et al. (2015).

**Internal validity and reliability**. One limitation is related to the results found since the classification of SE challenges and practices was performed via a human process. The identified challenges were classified based on a thematic analysis (Cruzes and Dybå, 2011), which also depends on the manual classification. The classification of the challenges followed the thematic analysis steps in order to identify the challenges reported in the studies that were faced by the professionals. In addition, the context in which each SE practice was applied was also classified based on the researchers' understanding of the context described in each paper. To reduce the bias in the classification of SE challenges and practices, this SLR involved three researchers who verified the classification of SE challenges and practices by resolving conflicts and regulating individual bias. Furthermore, the classification of SE practices was based on the SWEBOK knowledge areas (Bourque, P., Fairley, 2014) and contribution types proposed by Berg et al. (2018). Finally, the entire challenges and practices coding process was reviewed by two researchers who reviewed and discussed the information generated after each stage.

**External Validity**. Another threat is not to obtain a set of studies that represent all software engineering research aimed at supporting the development of AI and ML systems. This threat was minimized by using two strategies: (i) applying the research to four widely known libraries, such as Scopus, ACM, IEEE and Engineering Village. Scopus is a meta library and indexes publications by several well-known and reputable publishers, such as ACM, IEEE, Springer, and Elsevier; and (ii) by following a snowballing strategy to obtain the best possible coverage of the relevant literature for this research. With the use of the snowballing strategy, we sought papers that presented challenges and SE practices applied in the development of AI/ML systems. The SLR was also performed iteratively, that is, by conducting pilot research and refining the search terms, and validating the set of inclusion and exclusion criteria during the process. Well-accepted systematic methods and processes have been applied throughout the study and have been documented in the research protocol so that this study can be replicated by other researchers who are interested in SE research applied in AI/ML systems.

## 6. Conclusion

In this study, we have performed a systematic literature review in order to analyze the literature related to software engineering that supports the development of AI/ML systems. A total of 57 primary papers were extracted and synthesized. Our study is among the first comprehensive reviews that look at contexts, challenges, and practices of developing AI/ML software in the period from 1990 to 2019.

The contribution of this SLR study is two-fold. Firstly, the study provides a comprehensive view of the software industry for the development of AI/ML systems, and lists the main challenges faced. Possible research gaps are derived from future studies. Secondly, the study provides a map of the contextual setting of investigated SE practices, inferring the applicability area of empirical findings. This can help when comparing and generalizing future research of SE practices for AI/ML systems.

To address the main research question *"How is software engineering applied in the development of AI/ML systems?"*, we organized our findings from the research questions: (RQ1) Type of AI/ML systems developed; (RQ2) Context of the organization in which the systems are developed; (RQ3) The challenges faced by software professionals when developing ML/AI systems; (RQ4) How SE practices are being used to support the development of these systems; (RQ5) The SE practices reported in the primary studies; and (RQ6) The experimental methods used.

Regarding the type of AI/ML (RQ1) systems developed, our result showed that there are a variety of domains for AI/ ML applications, among these, the most prominent were automotive, health and finance. The most used paradigm in the development of these systems is supervised learning and this has a significant number of applications using neural network/deep learning approaches. In most cases in this research, the context (RQ2) reports AI development as a research-driven process. Significant AI projects are reported in a lab context or a large company context. However, there is a lack of process studies in SME or start-up contexts.

We list the challenges (RQ3) evidenced in the literature in a single list, and we performed a thematic analysis (Cruzes and Dybå, 2011) of the citations referenced in the papers regarding the challenges faced in the development of AI/ML. Thirteen categories of challenges were identified, and five categories were highlighted: Test, Software Quality, Data Management, Model Development, and Project Management.

We found that SE practices are being used (RQ4) mainly in the laboratory, mainly in the test area, and in the software industry, and applied more often in large companies by teams specialized in ML. Additionally, we list the SE practices (RQ5) and classify them by SWEBOK knowledge areas and the type of contribution. The areas of testing, configuration management, design, and construction were the ones that had the highest number of proposals. The majority of contribution types are for the framework, guidelines, lessons learned, and tools.

Based on our results, we have built a knowledge base on the challenges and practices in software engineering, and have reported some important insights regarding different contexts, as well as highlighting the challenges, practices, and types of contributions that we believe to be useful for researchers and professionals working in the field of AI/ML. For example, software professionals can find a list of practices that can be applied to the context of their project, such as applying AI ethics strategies, continuous and automated integration to the data pipeline, among others.

Future work will focus on certain research topics, as there are still gaps in SE practices in relation to some areas of knowledge such as management, maintenance, and software quality. Furthermore, more work should be conducted for specific business contexts, such as start-ups and small companies that are part of incubators. As a next step, we intend to evaluate the practices found in this SLR study together with software professionals who work in the development of AI/ML projects. We also intend to verify that the proposed practices meet the identified challenges. Finally, we plan to create an open repository with a list of summarized and actionable of SE practices for AI/ML systems development. In this way, it will be possible for professionals and researchers to consult the information about practices and check the context in which the practices were applied, their type of contribution, the challenges, and associated SWEBOK knowledge areas.

## 7. Acknowledgments

We would like to thank the financial support granted by CNPq via process number 311494/2017-0. CAPES - PROAP 001, and FAPEAM via process numbers 062.00150/2020, and 002/2018 POSGRAD 2018 – UFAM. This research, according to Article 48 of Decree nº 6.008/2006, was partially funded by Samsung Electronics of Amazonia Ltda, under the terms of Federal Law nº 8.387/1991, through agreement nº 003/2019, signed by ICOMP/UFAM. We special thanks to the researcher Edson César for the support and collaboration during the conduct of this study.

**Appendix A**

The appendix presents the list complete the primary studies selected in the 2nd filter (A.17) and shows the research method and research strategy (RQ6) applied in each study.

**Table A.17**

List of the 55 primary studies selected for SLR.

| Paper ID | Title | Authors | Research method | Research strategy |
|---|---|---|---|---|
| **P1** | Understanding Development Process of Machine Learning Systems: Challenges and Solutions | Nascimento et al. (2019) | Case study | Field study, Judgement study |
| **P2** | Software Engineering for Machine Learning: A Case Study | Amershi et al. (2019) | Case study | Field Study, Judgement study |
| **P3** | How do Engineers Perceive Difficulties in Engineering of Machine-Learning Systems? - Questionnaire Survey | Ishikawa (2019) | Survey | Judgement study |
| **P4** | Implementing Ethics in AI: Initial results of an industrial multiple case study | Vakkuri et al. (2019) | Case study | Field study |
| **P5** | A Taxonomy of Software Engineering Challenges for Machine Learning Systems: An Empirical Investigation | Lwakatare et al. (2019) | Case study | Field study |
| **P6** | Implementing AI ethics in practice: An empirical evaluation of the RESOLVEDD strategy | Vakkuri and Kemell (2019) | Case study | Field study |
| **P7** | Taming Functional Deficiencies of Automated Driving Systems: A Methodology Framework toward Safety Validation | Chen et al. (2018) | Case study | Field study, Field experiment |
| **P8** | Your System Gets Better Every Day You Use It: Towards Automated Continuous Experimentation | Mattos et al. (2017) | Case study | Field study, Sample study |
| **P9** | DataLab: Introducing software engineering thinking into data science education at scale | Zhang et al. (2017) | Case study, survey | Field study, Judgement study |
| **P10** | Trials and tribulations of developers of intelligent systems: A field study | Hilllaz et al. (2016) | Survey | Judgement study |
| **P11** | Lean Data Science Research Life Cycle: A Concept for Data Analysis Software Development | Shcherbakov et al. (2014) | Case study | Field experiment |
| **P12** | An empirical study of bugs in machine learning systems | Thung et al. (2012) | Experiment | Lab experiment |
| **P13** | Beyond the prototype: The design evolution of a deployed AI system | Martin and Schreckenghost (2004) | Case study | Field study |
| **P14** | Software development for CIM - a | Kouba et al. | Case study | Field study |
| | case study | (1992) | | |
| **P15** | An Empirical Study towards Characterizing Deep Learning Development and Deployment across Different Frameworks and Platforms | Guo et al. (2019) | Case study | Field experiment |
| **P16** | Reusability in Artificial Neural Networks: An Empirical Study | Ghofrani et al. (2019) | Survey | Judgement study |
| **P17** | Metric-Based Evaluation of Multiagent Systems Models | Damasceno et al. (2018) | Case study | Field study |
| **P18** | Guiding Deep Learning System Testing Using Surprise Adequacy | Kim et al. (2019) | Case study | Field experiment |
| **P19** | Requirements and Initial Model for KnowLang: A Language for Knowledge Representation in Autonomic Service-Component Ensembles | Vassev et al. (2011) | Case study | Field study |
| **P20** | Black Box Fairness Testing of Machine Learning Models | Aggarwal et al. (2019) | Experiment | Lab experiment |
| **P21** | Bridging the Gap between ML Solutions and Their Business Requirements Using Feature Interactions | Barash et al. (2019) | Experiment | Field experiment |
| **P22** | Boosting Operational DNN Testing Efficiency through Conditioning | Li et al. (2019) | Experiment | Lab experiment |
| **P23** | Stress Testing an AI-Based Web Service: A Case Study | Chakravarty (2010) | Case study | Field experiment |
| **P24** | The Emerging Role of Data Scientists on Software Development Teams | Kim et al. (2016) | Survey | Judgement study |
| **P25** | The current status of expert system development and related technologies in Japan | Motoda (1990) | Survey | Judgement study |
| **P26** | On the Engineering of AI-Powered Systems | Kusmenko et al. (2019) | Case study | Field experiment, Simulation |
| **P27** | Why is Developing Machine Learning Applications Challenging? A Study on Stack Overflow Posts | Alshangiti et al. (2019) | Survey | Sample study |
| **P28** | A Practical Framework for Agent-Based Hybrid Intelligent Systems | Li and Li (2011) | Case study | Field experiment |
| **P29** | Studying Software Engineering Patterns for Designing Machine Learning Systems | Washizaki et al. (2019) | Survey | Sample study, Judgement study |
| **P30** | Data Management Challenges for Deep Learning | Munappy et al. (2019) | Case study | Field experiment |
| **P31** | Continuous integration of machine learning models with ease.ml/ci: Towards a rigorous yet practical treatment | Renggli et al. (2019) | Experiment | Lab experiment |
| **P32** | Software engineering challenges of deep learning | Arpteg et al. (2018) | Case study | Field study |
| **P33** | Hidden technical debt in machine learning systems | Sculley et al. (2015) | Case study | Judgement study |
| **P34** | Developing Bug-Free Machine Learning Systems with Formal Mathematics | Selsam et al. (2017) | Experiment | Lab experiment |
| **P35** | Machine learning system architectural pattern for improving operational stability | Yokoyama (2019) | Experiment | Lab experiment |
| **P36** | On Testing Machine Learning Programs. | Braiek and Khomh (2020) | Survey | Sample study |
| **P37** | The challenge of verification and testing of machine learning | Goodfellow and Papernot (2017) | Experiment | Lab experiment |
| **P38** | Standardizing Ethical Design for Artificial Intelligence and Autonomous Systems | Bryson and Winfield (2017) | Experiment | Lab experiment |
| **P39** | How do scientists develop and use scientific software? | Hannay et al. (2009) | Survey | Sample study |
| **P40** | Validating a deep learning framework by metamorphic testing | DIng et al. (2017) | Experiment | Lab experiment |
| **P41** | Data Scientists in Software Teams: State of the Art and Challenges | Kim et al. (2018) | Survey | Field study |
| **P42** | Engineering safety in machine learning | Varshney (2017) | Case study | Field study |
| **P43** | DeepXplore: Automated Whitebox Testing of Deep Learning Systems | Pei et al. (2019) | Experiment | Lab experiment |
| **P44** | DeepGauge: multi-granularity testing criteria for deep learning systems | L. Ma et al. (2018) | Experiment | Lab experiment |
| **P45** | Machine learning software engineering in practice: An industrial case study | Rahman et al. (2019) | Case study | Field study |
| **P46** | Rules of machine learning: Best practices for ML engineering | Martin (2016) | Survey | Field study, Judgement study |
| **P47** | Feature-Guided Black-Box Safety Testing of Deep Neural Networks | Wicker et al. (2018) | Experiment | Lab experiment |
| **P48** | Deeproad: Gan-based metamorphic testing and input validation framework for autonomous driving systems | Zhang et al. (2018) | Experiment | Lab experiment |
| **P49** | An architectural framework for the construction of hybrid intelligent forecasting systems: application for electricity demand prediction | Lertpalangsunti and Chan (1998) | Experiment | Field experiment |
| **P50** | Solution patterns for machine learning | Nalchigar et al. (2019) | Experiment | Field experiment |
| **P51** | MODE: automated neural network | S. Ma et al. | Experiment | Lab |
| | model debugging via state differential analysis and input selection | (2018) | | experiment |
| **P52** | Testing and validating machine learning classifiers by metamorphic testing | Xie et al. (2011) | Case study | Field study |
| **P53** | Concolic Testing for Deep Neural Networks. | Sun et al. (2018b) | Experiment | Lab experiment |
| **P54** | Testing Deep Neural Networks | Sun et al. (2018a) | Experiment | Lab experiment |
| **P55** | DeepTest: Automated Testing of Deep-neural-network-driven Autonomous Cars | Jana et al. (2018) | Experiment | Field experiment, Lab experiment |

**Appendix B**

The appendix summarizing the categories and subcategories of the challenges found in the development of AI/ML systems. Along with the total number of papers that reference the challenge, the code of the paper references the challenge and the total number of challenges citations in the paper.

**Table B.18**

Categories and subcategories of the challenges (1990-2019), based on Fig. 10.

| # | Category and subcategories | Total Files | % | Papers | Total References |
|---|---|---|---|---|---|
| **1** | **Testing** | 25 | **44%** | P15, P18, P19, P21, P22, P3, P31, P33, P38, P44, P45, P46, P49, P53, P54 | 30 |
| | Bug detection | 2 | | P36, P39 | 2 |
| | Create test cases | 1 | | P22 | 1 |
| | Data testing | 1 | | P29 | 1 |
| | Debugging | 1 | | P32 | 1 |
| | Lack of oracle | 4 | | P18, P36, P44, P53 | 4 |
| | Test criterion | 1 | | P54 | 1 |
| **2** | **AI Software Quality** | 26 | **46%** | P5, P32 | 27 |
| | Interpretability | 5 | | P5, P32, P36 | 5 |
| | AI ethics requirement | 3 | | P4, P6, P38 | 4 |
| | Complexity | 1 | | P55 | 1 |
| | Efficiency | 1 | | P13 | 1 |
| | Fairness | 1 | | P19 | 1 |
| | Imperfection | 2 | | P18, P24 | 2 |
| | Data testing | 1 | | P32 | 1 |
| | Robustness | 2 | | P15, P53 | 2 |
| | Safety and Stability | 1 | | P3 | 1 |
| | Scalability | 3 | | P2, P40, P41 | 3 |
| | Stability | 1 | | P35 | 1 |
| | Staleness | 1 | | P35 | 1 |
| **3** | **Data Management** | 16 | **26%** | P2, P26, P29 | 16 |
| | Collecting data | 2 | | P5, P29 | 2 |
| | Data availability | 1 | | P41 | 1 |
| | Data dependency | 1 | | P33 | 1 |
| | Data quality | 1 | | P41 | 1 |
| | Handling data drift | 2 | | P5, P29 | 2 |
| | Integrating data | 1 | | P41 | 1 |
| | Slicing data | 1 | | P26 | 1 |
| | Structure of input data | 1 | | P1 | 1 |
| | Training data | 3 | | P3, P36, P40 | 3 |
| **4** | **Model development** | 16 | **28%** | P2, P15, P18, P26, P27 | 16 |
| | A large number of model inputs | 2 | | P18, P31 | 2 |
| | AI Ethics Implementation | 1 | | P4 | 1 |
| | Complex model formal | 1 | | P36 | 1 |

| # | Category and subcategories | Total Files | % | Papers | Total References |
|---|---|---|---|---|---|
| | representation | | | | |
| | Feature engineering optimization | 1 | | P41 | 1 |
| | Imperfection and Accuracy Assurance | 1 | | P3 | 1 |
| | Invalidation of models | 1 | | P5 | 1 |
| | Model localization with data constraints | 1 | | P5 | 1 |
| | Module documentation | 1 | | P16 | 1 |
| | Uncertainty in Input-Output relationships | 1 | | P18 | 1 |
| | Uncertainty in model behaviour | 1 | | P18 | 1 |
| **5** | **Project Management** | 15 | **26%** | P3, P11 | 16 |
| | Communication | 2 | | P6, P18 | 2 |
| | Competence | 4 | | P13, P18, P26, P34 | 5 |
| | Cultural factors | 1 | | P32 | 1 |
| | Estimation | 1 | | P32 | 1 |
| | Feedback | 1 | | P11 | 1 |
| | New node (experiments with real data) | 1 | | P8 | 1 |
| | Organizational factors | 1 | | P2 | 1 |
| | Process management | 1 | | P33 | 1 |
| | Risk management | 1 | | P7 | 1 |
| **6** | **Infrastructure** | 14 | **25%** | P2, P5, P13, P26, P32, P36 | 17 |
| | Configuration debt | 1 | | P33 | 1 |
| | Data management tools | 1 | | P41 | 1 |
| | Logging mechanisms | 2 | | P5, P10 | 2 |
| | Platform | 1 | | P32 | 1 |
| | Resource limitation | 1 | | P32 | 1 |
| | Supporting tools | 2 | | P6, P10 | 2 |
| **7** | **Requirement Engineering** | 13 | **23%** | P2, P16, P45, P50 | 15 |
| | Business Metric | 1 | | P1 | 1 |
| | Conflicting requirements | 1 | | P8 | 1 |
| | Insufficient understanding of customers | 1 | | P3 | 1 |
| | No method for requirement validation | 1 | | P41 | 1 |
| | Real user data | 1 | | P23 | 1 |
| | Specification | 2 | | P19, 50 | 4 |
| | Too much expectation | 1 | | P3 | 1 |
| | Undeclared customers | 1 | | P32 | 1 |
| **8** | **AI Engineering** | 10 | **18%** | P5 | 11 |
| | Automation | 1 | | P8 | 2 |
| | Combining different processes | 2 | | P10, P25 | 2 |

| # | Category and subcategories | Total Files | % | Papers | Total References |
|---|---|---|---|---|---|
| | Defined development process | 3 | | P1, P8, P18 | 3 |
| | Lack of patterns | 1 | | P28 | 1 |
| | Need for continuous engineering | 2 | | P3, P5 | 2 |
| **9** | **Architecture Design** | 8 | **14%** | P3, P5, P45 | 8 |
| | Architecture Evaluation | 1 | | P8 | 1 |
| | Changing anything, changing everything | 1 | | P33 | 1 |
| | Hard for strict abstraction boundaries | 1 | | P33 | 1 |
| | ML antipatterns | 1 | | P33 | 1 |
| | Risk factors for design | 1 | | P33 | 1 |
| **10** | **Model deployment** | 6 | **11%** | P15, P26 | 6 |
| | Dependency management | 1 | | P32 | 1 |
| | Maintaining glue code | 1 | | P32 | 1 |
| | Monitoring and logging | 1 | | P32 | 1 |
| | Unintended feedback loops | 1 | | P32 | 1 |
| **11** | **Integration** | 2 | **4%** | P2, P14 | 2 |
| **12** | **Operation Support** | 1 | **2%** | P2 | 2 |
| **13** | **Education** | 1 | **2%** | P9 | 1 |

**Appendix C**

The appendix presents SE Practices for AI/ML grouped into by the SWEBOK knowledge area, together with the applied context, contribution types, and literature references.

**Table C.19**

List of the 51 SE Practices for AI/ML systems (1990-2019).

| ID | Practice | Description | Authors | SWEBOK Knowledge Areas | Contribution types | Context |
|---|---|---|---|---|---|---|
| **P1** | Checklists for ML systems development process support: business modelling and data processing | Checklist to support Business Modelling (CheckBM) - created a set of verification criteria for each task performed by the developer at the initial stage of the project.<br>Checklist to support in Data Processing (CheckDP) - developed to support the structuring of the data task. In this stage, the developer performs different subtasks (e.g., verifying that the data is complete). | Nascimento et al. (2019) | Requirements, Design, Process | Framework/ methods | Software companies with few specialized ML teams. In some cases, with novice engineers. |
| **P2** | Lessons learned from the IA/ML projects at Microsoft - The best practices set | Viewpoints on some of the essential challenges associated with building large- scale ML applications and platforms and how the teams In addition, are points highlighting what is essential to the practice of AI in software teams. | Amershi et al. (2019) | Design, Construction, Configuration Management, Deployment, Maintenance, Professional Practice | Lessons learned | Large software companies with strong, specialized ML teams |
| **P3** | Lesson learning for difficulties in the engineering of ML-based systems | Presents a set lesson learned on how practitioners currently perceive difficulties and their causes in the engineering of ML-based systems, such as (i) decision making with customers; (ii) testing & QA; and (iii) Machine | Ishikawa (2019) | Requirements, Testing Maintenance | Lessons learned | Survey with participants that work with ML systems in various domains in the industry (Japan). |
| | | Learning systems engineering. | | | | |
| **P4** | Guidelines for implementing ethics in AI | Guidelines are presented regarding aspects of the ethics in the development process, such as ethics by design (integration of ethical reasoning capabilities as a part of system behaviour e.g. ethical robots), ethics in design (the regulatory and engineering methods), and ethics for design (codes of conduct, standards, etc.). | Vakkuri et al. (2019) | Design, Professional Practice | Guidelines | Large software companies with few AI specialists |
| **P5** | Practices to Automated Driving Software development | Practices to interpret sensor data to understand information contained at a high level. | Lwakatare et al. (2019) | Configuration Management | Lessons learned | Large software companies with strong, specialized ML teams |
| **P5** | An AI platform for support in data science teams | An AI platform was created to communicate requirements to the product team, provide internal AI education, and do projects with external companies. | Lwakatare et al. (2019) | Configuration Management | Lessons learned | Large software company with a strong data science team |
| **P5** | Federated learning for predicting failures in Mobile Network Operations | In extreme scenarios with a dataset of 3TB per day and where data is not allowed to be moved outside a country, federated learning is used. | Lwakatare et al. (2019) | Configuration Management | Lessons learned | Large software companies with strong, specialized ML teams |
| **P5** | A Platform to automating information extraction to optimize communication between team and stakeholders | A Platform to labeled and form the validation dataset for building ML models and automating information extraction of the databases. | Lwakatare et al. (2019) | Configuration Management | Lessons learned | Large software companies with a strong data science team |
| **P5** | Online Experimentation Platform | The experimentation platform consists of mainly four components namely the experimentation portal, | Lwakatare et al. (2019) | Configuration Management | Tool | Large software companies with a strong data |
| | | experiment execution service, log processing service, and analysis service. | | | | science team |
| **P6** | A framework to support AI Ethics - RESOLVEDD Strategy | Framework RESOLVEDD strategy considers ethical aspects, such as commitment to ethically aligned design, transparency in design, accountability in design, and responsibility in design. | Vakkuri and Kemell (2019) | Design, Process, Professional Practice, Configuration Management | Framework/methods | Academic Context |
| **P7** | A Methodology Framework toward Safety Validation for automated driving systems | A methodology based in Causal Scenario Analysis (CSA) has been introduced as a novel systematic paradigm to derive and evaluate the scenarios in automated driving systems. | Chen et al. (2018) | Requirements, Design, Process, Configuration Management | Framework/methods | NA |
| **P8** | An architecture framework to provide Automated Continuous Experimentation | The framework supports automated experimentation to ensure architectural qualities such as external adaptation control, data collection, performance, explicit representation of the learning component, decentralized adaptation, and knowledge exchange. | Mattos et al. (2017) | Design, Configuration Management | Framework/methods | Large software company with strong, specialized R&D team ML |
| **P9** | DataLab a framework to Data Science Education | The DataLab is a platform to help to improve the data science education in the classroom. | Zhang et al. (2017) | Professional Practice | Framework/methods | Academic Context |
| **P10** | Practical advice for ML-specialized software engineering | A few possible directions forward for software engineering practice of intelligent systems development, such as artifacts use, tools, and technical debt. | Hilllaz et al. (2016) | Process, Testing, Configuration Management, Professional Practice | Advice/implications | Survey with participants that work with ML systems in various domains in the industry |
| **P11** | Applying the lean data science research | This approach permits considering data science research in the framework of the | Shcherbakov et al. (2014) | Process | Theory | NA |
| | lifecycle based on the key principles of lean development | decision support process. | | | | |
| **P12** | Lessons learned to investigate bugs in ML systems | The authors use the same way to investigate software bugs for ML bugs and lessons learned were created based on the identification of ML systems bugs. | Thung et al. (2012) | Testing | Lessons learned | NA |
| **P13** | Lessons learned about object-oriented principles of encapsulation and polymorphism to AI-based systems | The summary of lessons learned traces back the general object-oriented principles of encapsulation and polymorphism as well as the well-recognized need for run-time exception handling AI-based systems. | Martin and Schreckenghost (2004) | Design, Construction, Configuration Management, Professional Practice | Lessons learned | Large software company with AI specialized teams. |
| **P14** | Model of the RUDE cycle (Run - Understand - Debug - Edit) to AI systems development | The model of the RUDE cycle is used to incremental improvements as a fast-prototyping strategy that is used in the AI systems development quite. | Kouba et al. (1992) | Requirement, Process | Model | Software company with strong AI specialists. |
| **P15** | Lessons learned for issues and bugs detection to evaluate new DL frameworks and platforms | Useful guidance for deep learning developers and researchers, such as deep learning software bug detection when model migrated and quantized under different deep learning platforms and model conversion | Guo et al. (2019) | Design, Construction, Testing, Configuration Management, Deployment | Lessons learned | NA |
| **P16** | Lessons learned to apply reusability methods in the development of ANN-based systems. | Lessons learned are spotlights on the state of reusability in the development of ANN-based projects | Ghofrani et al. (2019) | Design, Construction, Configuration Management | Lessons learned | Survey with participants that work with ML systems in various domains in the industry |
| **P17** | Guidelines of Metric-Based Evaluation of | This proposal includes quality dimensions for multiagent systems | Damasceno et al. | Requirements, Design, | Guidelines | Academic Context |
| | Multiagent Systems Models | (mas), such as model expectations, resources quality, reliability, methodology adequacy, maintainability, and agent quality. | (2018) | Configuration Management, Quality, Maintenance | | |
| **P19** | A Model for KnowLang: A Language for Knowledge Representation in Autonomic Service-Component Ensembles | Every ASCENS Knowledge Corpus is structured into a domain-specific ontology, logical framework, and inter-ontology operators. The logical framework provides additional to the domain ontology computational structures that determine the logical foundations helping a Service Component (SC) reason and infer knowledge. | Vassev et al. (2011) | Requirement, Design, Configuration Management | Model | NA |
| **P20** | A methodology for auto-generation of test inputs | A methodology for auto-generation of test inputs, for the task of detecting individual discrimination. | Aggarwal et al. (2019) | Testing | Framework/ methods | NA |
| **P21** | A combinatorial modelling to gain insights on the quality of ML solutions | Combinatory testing for AI with data-based, defining tests data subsets or slices over the ML solution data. A combinatorial modeling methodology is a viable approach for gaining actionable insights on the quality of ML solutions. Applying the approach results in identifying coverage gaps indicating uncovered requirements. | Barash et al. (2019) | Testing | Model | NA |
| **P22** | DNN testing method to determine the trained DNN model actual performance | DNN testing method, i.e., testing a previously trained DNN model with the data collected from a specific operation context to determine the model's actual performance in this context. | Li et al. (2019) | Testing | Tool | NA |
| **P23** | An approach for stress testing that measuring performance in AI-Based Web Services | Stress testing that aims at measuring end-to-end performance. This practice proposes an efficient DNN testing method based on the conditioning on the representation learned by the DNN model under testing. The representation is defined by the probability distribution of the output of neurons in the last hidden layer of the model. | Chakravarty (2010) | Testing | Tool | Large software company with specialized AI teams. |
| **P25** | Guiding to development of an expert system (AI) | Advice on building an expert system, such as (1) formalizing the problem, based on the classification tree, (2) providing search strategies, (3) generating a task-specific expert system (4) providing domain-specific knowledge based on the data structure. | Motoda (1990) | Requirement, Process | Advice/ implications | NA |
| **P26** | Language framework to model and train deep neural networks (DNN) | EMADL (EmbeddedMontiArcDL) is a language family combining component-based architecture with deep learning modeling to model and train deep neural networks (DNN). EMADL was constructed using the language composition principles language extension, language aggregation, and language embedding of the language workbench MontiCore 5. | Kusmenko et al. (2019) | Design, Construction | Framework/ methods | NA |
| **P28** | A Framework for Agent-Based Hybrid Intelligent Systems (HIS) | An architectural framework that overcomes the technical impediments and facilitates hybrid intelligent system | Li and Li (2011) | Design | Framework/ methods | Large software company with AI specialists |
| | | construction is built. The framework mainly consists of three parts: decision making, knowledge discovering, and distributed heterogeneous data resources (DHDR). | | | | |
| **P29** | A Model of Architectural and design (anti-patterns) for Machine Learning Systems | SE patterns presented for ML systems can be divided along two main dimensions: ML pipeline and SE development process. A model of architecture patterns was proposed to ML systems with patterns that apply to many stages of the pipeline or many phases of the development process. | Washizaki et al. (2019) | Design | Model | Survey with participants that work with ML systems in various domains in the industry |
| **P30** | Lessons learned to foresee roadblocks that may encounter while managing data for deep learning systems | A set of 20 challenges were identified and lessons learned are described across the phases of the data pipeline. | Munappy et al. (2019) | Testing | Lessons learned | Large software companies with a strong data science team |
| **P31** | Ease.ml/ci - a continuous integration system for ML | Ease.ml/ci, is a continuous integration system for machine learning. It provides a declarative scripting language that allows users to state a rich class of test conditions with rigorous probabilistic guarantees. | Renggli et al. (2019) | Testing, Configuration Management | Tool | Large software company with strong, specialized ML teams |
| **P32** | Map of SE challenges to building systems with DL components | A set of 12 challenges were identified and outline main SE challenges with building systems with DL components that were described in the areas of development, production, and organizational | Arpteg et al. (2018) | Design, Construction, Configuration Management, Professional Practice | Theory | Large software company with strong, specialized ML teams |
| | | challenges. | | | | |
| **P33** | A framework of anti-patterns to technical debt in ML systems | A variety of is explored several ML-specific risk factors for in system design, included anti-patterns. A set of anti-patterns and practices to avoid technical debt in systems using ML components are presented. | Sculley et al. (2015) | Design, Construction, Maintenance | Framework/ methods | Large software company with specialized ML teams |
| **P34** | An interactive proof assistant for support in the implementation of the ML system | An interactive proof assistant to both implement their ML system and to state a formal theorem defining what it means for their ML system to be correct was proposed. | Selsam et al. (2017) | Testing, Configuration Management | Tool | NA |
| **P35** | An architectural pattern to improves the operational stability of ML systems | A novel architectural pattern that improves the operational stability of machine learning systems. This architectural pattern helps the operators break down the failures into a business logic part and an ML-specific part. | Yokoyama (2019) | Design, Configuration Management | Model | NA |
| **P37** | CleverHans - Reproducible testing for ensuring the robustness of a model | The CleverHans library contains reference implementations of several attack and defense procedures to test ML models against standardized, state-of-the-art attacks and defenses | Goodfellow and Papernot (2017) | Testing | Tool | NA |
| **P38** | ETHICS IN AI - Ethical Considerations to Artificial Intelligence and Autonomous Systems | Standards are consensus-based agreed-upon ways of doing things, setting out how things should be done, provides confidence in a system's efficacy in areas important to users, such as safety, security, and reliability. | Bryson and Winfield (2017) | Design, Construction | Advice/ implications | NA |
| **P39** | Guidelines of Software | Provides a list of practices and guidelines | Hannay et al. | Testing, Configurat | Guidelines | Survey with scientists |
| | Engineering Practices for data scientists | that can be used in the work of the data scientist. | (2009) | ion Managem ent, Maintenan ce, Profession al Practice | | that work with data science in various domains in the industry |
| **P40** | Metamorphic testing - validation approach of the deep learning framework for automated classifying biology cell images | An approach for validating the classification accuracy of a deep learning framework that includes a convolutional neural network, a deep learning executing environment, and a massive image data set. | DIng et al. (2017) | Testing | Frame work/ method s | NA |
| **P41** | Best practices to improve and ensure the quality of data scientists work | The best practices to overcome the challenge of data science work and advice that data scientists "expert" would give to novice data scientists. | Kim et al. (2018) | Requireme nt, Design, Constructi on, Testing, Process, Configurat ion Managem ent, Profession al Practice | Lessons learned | Survey with participants that work with data science or machine learning in various domains at Microsoft |
| **P42** | Four different strategies for achieving safety to the machine learning context | Strategies for achieving safety in ML projects to definite safety in terms of risk, epistemic uncertainty, and the harm incurred by unwanted outcomes. | Varshn ey (2017) | Design, Constructi on, Configurat ion Managem ent, Profession al Practice | Guideli nes | NA |
| **P43** | DeepXplore - Whitebox testing framework for large-scale DL systems | DeepXplore formulates the problem of generating test inputs that maximize neuron coverage of a DL system while also exposing as many differential behaviour (i.e., differences between multiple similar DL systems) as a joint | Pei et al. (2019) | Testing | Tool | NA |
| | | optimization problem. | | | | |
| **P44** | DeepGauge - a set of multi-granularity testing criteria for DL systems | Guidelines for researchers and practitioners with best practices for the development of ML applications, with three distinct perspectives: software engineering, machine learning, and industry-academia collaboration. | L. Ma et al. (2018) | Testing | Tool | NA |
| **P45** | Guidelines for researchers and practitioners for best practices in the development of ML applications. | A set of 43 best practices for machine learning from Google. That best practices serve as Guides together with other popular guides to practical programming. | Rahman et al. (2019) | Requirement, Design, Construction, Testing, Process, Configuration Management, Professional Practice | Guidelines | Large software companies with strong, specialized ML teams |
| **P46** | Guidelines of the best practices for machine learning | A feature-guided approach to test the resilience of image classifier networks against adversarial examples, with a novel feature-guided black-box algorithm. | Martin, (2016) | Requirement, Design, Construction, Testing, Process, Configuration Management, Professional Practice | Guidelines | Large software companies with strong, specialized ML teams |
| **P47** | A feature-guided black-box approach to test the safety of deep neural networks (DNN) | A feature-guided approach to test the resilience of image classifier networks against adversarial examples, with a novel feature-guided black-box algorithm | Wicker et al. (2018) | Testing | Tool | NA |
| **P48** | DeepRoad - to systematically analyse DNN-based autonomous driving | An unsupervised learning framework to systematically analyse autonomous DNN-based steering systems. DeepRoad consists of a | Zhang et al. (2018) | Testing | Tool | NA |
| | systems | metamorphic test module and an input validation module. | | | | |
| **P49** | An architectural framework to facilitate the design and implementation of hybrid intelligent forecasting applications | An architectural framework to facilitate the design and implementation of hybrid intelligent forecasting applications. The framework has been designed based on the authors' understanding of the subtasks necessary for performing forecasting. | Lertpalangsunti and Chan (1998) | Design, Construction | Framework/methods | Large software company with AI specialists |
| **P50** | A solution patterns to representing generic and well-proven ML designs for commonly-known and recurring business analytics problems | Introduces solution patterns as an explicit way of representing generic and well-proven ML designs for commonly-known and recurring business analytics problems | Nalchigar et al. (2019) | Design, Construction | Framework/methods | Large software company with specialists with few ML skills |
| **P51** | MODE - model debugging technique for neural network models | An automated neural network debugging technique powered by state differential analysis and input selection | S. Ma et al. (2018) | Design, Construction | Tool | NA |
| **P52** | A technique for testing the implementations of machine learning classification algorithms | A technique for testing the implementation of machine learning classification algorithms that support such applications, and is based on the technique "metamorphic testing" | Xie et al. (2011) | Testing, Construction | Framework/methods | NA |
| **P53** | DeepConcolic - a concolic testing approach for DNNs | DeepConcolic is generic and can take coverage requirements as input for testing for DNNs, and is the first concolic testing approach for DNNs. | Sun et al. (2018b) | Testing | Tool | NA |
| **P54** | White-box testing Framework for | A novel, white-box testing methodology for DNNs, with four test | Sun et al. (2018a) | Testing, Quality | Framework/method | NA |
| | DNNs based in the MC/DC test criterion | criteria, inspired by the MC/DC test criterion from traditional software testing, that fit the distinct properties of DNNs. | | | s | |
| **P55** | DeepTest - to systematically explore different parts of the DNN logic by generating test inputs that maximize the numbers of activated neurons | A systematic testing tool for automatically detecting erroneous behaviour of Deep Neural Networks (DNNs) DNN-driven vehicles. | Jana et al. (2018) | Testing | Tool | NA |

**Appendix D**

**Table D.20**

Results of quality assessment

| | | | | | | | | | | | |
|---|---|---|---|---|---|---|---|---|---|---|---|
| Score: **Yes = 1; Partially (Par) = 0.5; No = 0** | | | | | | | | | | | |
| **Score to paper quality: 7 – 10: high rigour; 4 – 6.5: medium rigour; 0 – 3.5: low rigour** | | | | | | | | | | | |
| ID | **Q1** | **Q2** | **Q3** | **Q4** | **Q5** | **Q6** | **Q7** | **Q8** | **Q9** | **Q10** | **Score** |
| P1 | Yes | Yes | Yes | Yes | Yes | Yes | Yes | Yes | Yes | Yes | **10.0** |
| P2 | Yes | Yes | Yes | Yes | Yes | Yes | Yes | Yes | Yes | Yes | **10.0** |
| P3 | Yes | Yes | Yes | Par | Par | Yes | Yes | No | Yes | Yes | **8.0** |
| P4 | Yes | Yes | Yes | Yes | Yes | Yes | Yes | Yes | Yes | Yes | **10.0** |
| P5 | Yes | Yes | Yes | Yes | Yes | Yes | Yes | Yes | Yes | Yes | **10.0** |
| P6 | Yes | Yes | Yes | Yes | Yes | Yes | Par | Yes | Yes | Yes | **9.5** |
| P7 | Yes | Yes | Yes | Yes | Yes | Yes | Yes | Yes | Yes | Yes | **10.0** |
| P8 | Yes | Yes | Yes | Yes | Yes | Yes | Par | Par | Yes | Yes | **9.0** |
| P9 | Yes | Yes | Yes | Yes | Yes | Yes | Par | Yes | Yes | Yes | **9.5** |
| P10 | Yes | Yes | Yes | Yes | Yes | Yes | Yes | Yes | Yes | Yes | **10.0** |
| P11 | Yes | Yes | Yes | Yes | No | Par | Par | No | Yes | Yes | **7.0** |
| P12 | Yes | Yes | Yes | Yes | Yes | Yes | Par | Yes | Yes | Yes | **9.5** |
| P13 | Yes | Yes | Yes | Yes | Par | Par | Par | Yes | Yes | Yes | **8.5** |
| P14 | Yes | Yes | Yes | Yes | Par | Par | Par | Yes | Yes | Yes | **8.5** |
| P15 | Yes | Yes | Yes | Yes | Yes | Yes | Yes | Yes | Yes | Yes | **10.0** |
| P16 | Yes | Yes | Yes | Yes | Yes | Yes | Yes | No | Yes | Yes | **9.0** |
| P17 | Yes | Yes | Yes | Yes | Yes | Yes | Yes | Yes | Yes | Yes | **10.0** |
| P18 | Yes | Yes | Yes | Yes | Yes | Yes | Yes | Yes | Yes | Yes | **10.0** |
| P19 | Yes | Yes | Yes | Yes | Yes | Yes | Yes | Yes | Yes | Yes | **10.0** |
| P20 | Yes | Yes | Yes | Yes | No | Par | Par | Yes | Yes | Yes | **8.0** |
| P21 | Yes | Yes | Yes | Yes | No | Yes | Yes | Yes | Yes | Yes | **9.0** |
| P22 | Yes | Yes | Yes | Yes | No | No | No | Yes | Yes | Yes | **7.0** |
| P23 | Yes | Yes | Par | No | No | Par | Par | Par | Par | Yes | **5.5** |
| P24 | Yes | Yes | Yes | Yes | No | Yes | Yes | Par | Yes | Yes | **8.5** |
| P25 | Par | Yes | Yes | No | Yes | Par | Par | No | No | Yes | **5.5** |
| P26 | Yes | Yes | Par | Yes | No | No | No | No | Yes | Yes | **4.5** |
| P27 | Yes | Yes | Yes | Yes | Yes | Yes | Yes | Yes | Yes | Yes | **10.0** |

| | | | | | | | | | | |
|---|---|---|---|---|---|---|---|---|---|---|
| P28 | Yes | Yes | Yes | No | No | No | No | No | No | Yes | **4.0** |
| P29 | Yes | Yes | Yes | Yes | Yes | Yes | Yes | Par | Yes | Yes | **9.5** |
| P30 | Yes | Yes | Yes | Yes | Par | Yes | Par | Par | Yes | Yes | **8.5** |
| P31 | Yes | Yes | Yes | Par | No | Par | Yes | Yes | Par | Yes | **7.0** |
| P32 | Yes | Yes | Yes | Yes | Yes | Yes | Yes | Yes | Yes | Yes | **10.0** |
| P33 | No | Yes | No | Yes | No | No | No | No | Yes | Yes | **4.0** |
| P34 | No | Yes | Yes | Par | No | No | No | No | Par | Yes | **4.0** |
| P35 | Yes | Yes | No | No | No | No | No | No | No | Yes | **3.0** |
| P36 | No | Yes | No | No | No | No | No | No | No | No | **1.0** |
| P37 | No | Yes | No | Yes | No | No | No | No | Yes | Yes | **4.0** |
| P38 | Par | Yes | No | Yes | Yes | Par | Par | No | Yes | Yes | **6.5** |
| P39 | Yes | Yes | Par | Par | No | Yes | Yes | No | Par | No | **4.5** |
| P40 | No | Yes | No | No | No | No | No | No | No | Yes | **2.0** |
| P41 | Yes | Yes | Yes | Par | Yes | Yes | Yes | No | Yes | Yes | **8.5** |
| P42 | No | Yes | No | Par | No | No | No | No | Yes | Yes | **3.5** |
| P43 | Yes | Yes | Yes | Yes | Yes | Yes | Par | Par | Yes | Yes | **9.0** |
| P44 | Yes | Yes | Yes | Yes | Yes | Yes | Yes | Yes | Yes | Yes | **10.0** |
| P45 | Yes | Yes | Yes | Yes | Yes | Yes | Yes | Yes | Yes | Yes | **10.0** |
| P46 | Yes | Yes | Par | Par | Yes | No | No | Yes | Yes | Yes | **7.5** |
| P47 | Yes | Yes | Yes | Yes | Yes | Yes | Yes | Yes | Yes | Yes | **10.0** |
| P48 | Yes | Yes | Yes | Yes | Yes | Yes | Yes | Yes | Yes | Yes | **10.0** |
| P49 | Yes | Yes | Yes | Yes | Yes | Yes | Yes | No | Par | Yes | **8.5** |
| P50 | Yes | Yes | Yes | Yes | Yes | Yes | Yes | Yes | Yes | Yes | **10.0** |
| P51 | Yes | Yes | Yes | Yes | Yes | Yes | Yes | No | Yes | Yes | **9.0** |
| P52 | Yes | Yes | Yes | Yes | Yes | Yes | Yes | No | Yes | Yes | **9.0** |
| P53 | Yes | Yes | Yes | Yes | Yes | Yes | Yes | No | Yes | Yes | **9.0** |
| P54 | Yes | Yes | Yes | Yes | Yes | Yes | Yes | No | Yes | Yes | **9.0** |
| P55 | Yes | Yes | Yes | Yes | Yes | Yes | Yes | No | Yes | Yes | **9.0** |